\documentstyle[12pt]{article}
\advance\hoffset by -7mm
\makeatletter
\@addtoreset{equation}{section}
\makeatother
\renewcommand{\theequation}{\thesection.\arabic{equation}}
\renewcommand{\title}[1]{\null\vspace{25mm}

\noindent{\Large{\bf #1}}\vspace{10mm}

\noindent {\large By }}
\newcommand{\authors}[1]{\noindent{\large #1}\vspace{3mm}

}
\newcommand{\address}[1]{\noindent #1\vspace{5mm}

}
\renewcommand{\abstract}[1]{\vspace{10mm}

\noindent{\small{\em Abstract.} #1}\vspace{2mm}

}

\begin{document}

\def\be{\begin{equation}}
\def\ee{\end{equation}}
\def\bea{\begin{eqnarray}}
\def\eea{\end{eqnarray}}
\def\bean{\begin{eqnarray*}}
\def\eean{\end{eqnarray*}}
\def\ba{\begin{array}} \def\ea{\end{array}}
\newcommand {\equ}[1] {(\ref{#1})}
\def\6{\partial} \def\a{\alpha} \def\b{\beta}
\def\g{\gamma} \def\d{\delta} \def\ve{\varepsilon}
\def\e{\epsilon}
\def\z{\zeta} \def\h{\eta} \def\th{\theta}
\def\vt{\vartheta} \def\k{\kappa} \def\l{\lambda}
\def\m{\mu} \def\n{\nu} \def\x{\xi} \def\p{\pi}
\def\r{\rho} \def\s{\sigma} \def\t{\tau}
\def\Ph{\phi} \def\ph{\varphi} \def\ps{\psi}
\def\o{\omega} \def\G{\Gamma} \def\D{\Delta}
\def\Th{\Theta} \def\L{\Lambda} \def\S{\Sigma}
\def\PH{\Phi} \def\Ps{\Psi} \def\O{\Omega}
\def\sm{\small} \def\la{\large} \def\La{\Large}
\def\LA{\LARGE} \def\hu{\huge} \def\Hu{\Huge}
\def\ti{\tilde} \def\wti{\widetilde}
\def\non{\nonumber\\}
\def\={\!\!\!&=&\!\!\!}
\def\+{\!\!\!&&\!\!\!+~}
\def\-{\!\!\!&&\!\!\!-~}
\def\id{\!\!\!&\equiv&\!\!\!}
\renewcommand{\AA}{{\cal A}}
\newcommand{\BB}{{\cal B}}
\newcommand{\CC}{{\cal C}}
\newcommand{\DD}{{\cal D}}
\newcommand{\EE}{{\cal E}}
\newcommand{\FF}{{\cal F}}
\newcommand{\GG}{{\cal G}}
\newcommand{\HH}{{\cal H}}
\newcommand{\II}{{\cal I}}
\newcommand{\JJ}{{\cal J}}
\newcommand{\KK}{{\cal K}}
\newcommand{\LL}{{\cal L}}
\newcommand{\MM}{{\cal M}}
\newcommand{\NN}{{\cal N}}
\newcommand{\OO}{{\cal O}}
\newcommand{\PP}{{\cal P}}
\newcommand{\QQ}{{\cal Q}}
\newcommand{\RR}{{\cal R}}
\newcommand{\SS}{{\cal S}}
\newcommand{\TT}{{\cal T}}
\newcommand{\UU}{{\cal U}}
\newcommand{\VV}{{\cal V}}
\newcommand{\WW}{{\cal W}}
\newcommand{\XX}{{\cal X}}
\newcommand{\YY}{{\cal Y}}
\newcommand{\ZZ}{{\cal Z}}
\newcommand{\journal}[4]{{\em #1~}#2\,(19#3)\,#4;}
\newcommand{\aihp}{\journal {Ann. Inst. Henri Poincar\'e}}
\newcommand{\hpa}{\journal {Helv. Phys. Acta}}
\newcommand{\sjpn}{\journal {Sov. J. Part. Nucl.}}
\newcommand{\ijmp}{\journal {Int. J. Mod. Phys.}}
\newcommand{\physu}{\journal {Physica (Utrecht)}}
\newcommand{\pr}{\journal {Phys. Rev.}}
\newcommand{\jetpl}{\journal {JETP Lett.}}
\newcommand{\prl}{\journal {Phys. Rev. Lett.}}
\newcommand{\jmp}{\journal {J. Math. Phys.}}
\newcommand{\rmp}{\journal {Rev. Mod. Phys.}}
\newcommand{\cmp}{\journal {Comm. Math. Phys.}}
\newcommand{\cqg}{\journal {Class. Quantum Grav.}}
\newcommand{\zp}{\journal {Z. Phys.}}
\newcommand{\np}{\journal {Nucl. Phys.}}
\newcommand{\pl}{\journal {Phys. Lett.}}
\newcommand{\mpl}{\journal {Mod. Phys. Lett.}}
\newcommand{\prep}{\journal {Phys. Reports}}
\newcommand{\prepsec}{\journal {Phys. Reports (Review Section
                                of Phys. Letters)}}
\newcommand{\ptp}{\journal {Progr. Theor. Phys.}}
\newcommand{\nc}{\journal {Nuovo Cim.}}
\newcommand{\app}{\journal {Acta Phys. Pol.}}
\newcommand{\apj}{\journal {Astrophys. Jour.}}
\newcommand{\apjl}{\journal {Astrophys. Jour. Lett.}}
\newcommand{\annp}{\journal {Ann. Phys. (N.Y.)}}
\newcommand{\anp}{\journal {Ann. of Phys.}}
\newcommand{\Nature}{{\em Nature}}
\newcommand{\PRD}{{\em Phys. Rev. D}}
\newcommand{\MNRAS}{{\em M. N. R. A. S.}}

\title{On the Algebraic Structure of Gravity with Torsion\\[2mm]
       including Weyl symmetry}
\authors{O. Moritsch\footnote{Work supported in part by the
         ``Fonds zur F\"orderung der Wissenschaftlichen Forschung''
         under Contract Grant Number P9116-PHY.}
         and M. Schweda}
\address{Institut f\"ur Theoretische Physik, Technische Universit\"at Wien\\
         Wiedner Hauptstra\ss e 8-10, A-1040 Wien (Austria)}
\noindent{REF. TUW 94-08}
\abstract{
The BRST transformations for gravity with torsion including
Weyl symmetry are discussed by using the so-called
Maurer-Cartan horizontality conditions.
Also the coupling of scalar matter fields to gravity is incorporated
in this analysis. With the help of
an operator $\d$ which allows to decompose the exterior space-time
derivative as a BRST commutator we solve the Wess-Zumino consistency
condition corresponding to invariant Lagrangians and anomalies
for the cases with and without Weyl symmetry.
}


\subsection*{Contents:}

\begin{tabular}{l}
1 Introduction      \\
2 Basic elements    \\
3 Maurer-Cartan horizontality conditions    \\
4 Descent equations and decomposition   \\
5 Some examples   \\
6 The geometrical meaning of the operator $\d$  \\
7 Conclusion    \\
Appendix A: Commutator relations  \\
Appendix B: Determinant of the vielbein and the $\ve$ tensor \\
References
\end{tabular}


\section{Introduction}

In the discussion of the unification of all fundamental
interactions, gauge field theories play a central role.
Electroweak theory and quantum chromodynamic (QCD) are
examples of Yang-Mills gauge theories~\cite{yangmills,glashow}
associated with non-abelian Lie groups. In that way gravity
is introduced as a gauge theory which is associated with local
Lorentz invariance~\cite{utiyama}.

The symmetry content of a field theoretic model
is usually described by Ward identities (WI) leading to
functional differential equations for the various Green's
functions generated by the corresponding generating
functionals~\cite{zuber,piguet}.
Sometimes, the transition from the classical to the
quantized level modifies these Ward identities by
non-trivial contributions (anomalies), expressing the fact,
that the original
symmetry of the classical model is broken at the quantum level.

An anomaly is usually defined as the gauge variation of the
connected vacuum functional in the presence of external gauge fields.
When an anomaly occurs, this variation does not vanish and the vacuum
functional is not gauge invariant.

The most famous anomaly is the Adler-Bell-Jackiw (ABJ)
anomaly~\cite{adler,bell,bard} which describes the breaking term in the
axial vector current divergence equation.
This anomaly is needed to discuss successfully the $\p^{0}\rightarrow 2\g$
decay.

In connection with conformal field theories of gravity, Weyl anomalies
are of great interest. It is well-known
that due to the existence of Weyl anomalies the Weyl symmetry, which is
valid at the classical level, is broken in the presence of
quantum corrections.
It is apparent that in the discussion
about the quantum conformal structure of a theory
one needs the identification of all the Weyl
anomalies~\cite{duff,tonin3} and Weyl
invariants in arbitrary space-time dimensions.

Therefore, in order to discuss anomalies one needs a tool for a
characterization. This may be achieved in a very compact manner
with the help of the Wess-Zumino (WZ) consitency condition~\cite{wess}
in the context of the Becchi-Rouet-Stora-Tyupin (BRST)
formalism~\cite{brs}.
This BRST scheme is an elegant and powerful instrument for the
consistent discussion of gauge symmetries in quantum field theory,
and in addition this concept is available for a large class of
gauge field models whose classical symmetries have an algebra
which closes.
In particular this BRST formalism allows to characterize the
classical action and possible anomalies as BRST invariant local
functionals of the basic fields.

In order to describe the general procedure applicable to any gauge
field model one starts with the one-particle-irreducible (1PI)
vertex functional given by
\be
\label{VERTEX-1PI}
\G(\Ph_{cl})=\sum_{n=2}^\infty\frac{1}{n!}\int dx_{1}...dx_{n}
\Ph_{cl}(x_{1})...\Ph_{cl}(x_{n})
\langle 0|T(\Ph(x_{1})...\Ph(x_{n}))|0\rangle^{1PI} \ ,
\ee
where $\langle 0|T(\Ph(x_{1})...\Ph(x_{n}))|0\rangle^{1PI}$ denotes
the vacuum expectation value of the quantum fields $\Ph(x)$.
The classical sources $\Ph_{cl}$ are test functions for the
functional \equ{VERTEX-1PI}.
In perturbation theory one can make a loop expansion
for $\G(\Ph_{cl})$, i.e. the vertex functional can be written
as a formal power series in $\hbar$:
\be
\label{LOOP-EXP}
\G(\Ph_{cl})=\sum_{n=0}^{\infty}\hbar^{n}\G^{(n)}(\Ph_{cl}) \ .
\ee
At the classical level, in the so-called tree approximation,
one gets
\be
\label{VERTEX}
\G^{(0)}(\Ph_{cl})=\sum_{n=2}^\infty\frac{1}{n!}\int dx_{1}...dx_{n}
\Ph_{cl}(x_{1})...\Ph_{cl}(x_{n})
\langle 0|T(\Ph(x_{1})...\Ph(x_{n}))|0\rangle^{1PI}_{tree}
=\G_{cl}(\Ph_{cl}) \ ,
\ee
where $\langle 0|T(\Ph(x_{1})...\Ph(x_{n}))|0\rangle^{1PI}_{tree}$
collects now all possible tree graphs, i.e. graphs without radiative
corrections. In this approximation the zero-loop order corresponds
to the classical action.
In order to simplify notation one substitutes $\Ph_{cl}\rightarrow\Ph$
in the functionals \equ{VERTEX-1PI} and \equ{VERTEX}.

For the discussion of the symmetry content of the field model one introduces
the Ward identity operator $W_{s}$ in its global form
\be
\label{WIO}
W_{s}=\int d^{N}\!x~\d_{s}\Ph(x)\frac{\d}{\d\Ph(x)} \ ,
\ee
belonging to an arbitrary infinitesimal symmetry transformation $\d_{s}\Ph(x)$
characterized by a local parametric function $\ve(x)$
\be
\label{ST}
\d_{s}\Ph(x)=\ve(x)\PP(\Ph) \ ,
\ee
whereby for general reason $\PP(\Ph)$ may be linear or non-linear in $\Ph$.
Applying the WI-operator \equ{WIO} to \equ{VERTEX} one gets, for the case
that \equ{ST} is a symmetry of the model, the following global WI:
\be
\label{WI}
W_{s}\G_{cl}(\Ph)=0 \ .
\ee
The corresponding local WI may be obtained from \equ{WI} by functional
differentiation with respect to $\ve(x)$.

The presence of radiative corrections is now governed by the renormalized
action principle~\cite{lowenstein,lam} leading in general to a modified WI
for the full vertex functional \equ{LOOP-EXP}
\be
\label{RAP}
W_{s}\G(\Ph)=\D \cdot \G(\Ph)=\hbar\D(\Ph)+\OO(\hbar^{2}) \ ,
\ee
where $\D$ is an integrated well-defined quantum insertion of definite
dimensions~\cite{lowenstein,lam} and $\D(\Ph)$ is an integrated local
polynomial in the fields and their derivatives.
For the search of anomalies it is enough to
limit ourselves to the one-loop order.

Following the general procedure for the discussion of non-invariant
counterterms an anomaly occurs if the correction $\D$ cannot be expressed
as a variation of the underlying symmetry and therefore it is not possible
to absorb $\D(\Ph)$ by an appropriate counterterm in order to get
\be
W_{s}\widehat{\G}(\Ph)=0 \ .
\ee
To clarify this point, we assume that $\D(\Ph)$ can be written as
\be
\label{ANOMAL}
\D(\Ph)=a(\Ph)+W_{s}b(\Ph) \ ,
\ee
where $a(\Ph)$ and $b(\Ph)$ are integrated local polynomials and
where $a(\Ph)$ cannot be expressed as $W_{s}\widehat{a}(\Ph)$ for any
integrated local polynomial $\widehat{a}(\Ph)$.
The second term in \equ{ANOMAL} can be absorbed in a redefined
$\widehat{\G}(\Ph)$, but the first term leads to an anomaly
\be
\label{BREAK}
W_{s}\widehat{\G}(\Ph)=W_{s}(\G(\Ph)-\hbar b(\Ph))=\hbar a(\Ph) \ .
\ee

The use of the BRST scheme demands now that the given symmetry is converted
into a BRST symmetry. This can be achieved by replacing the infinitesimal
parametric functions $\ve(x)$ by anticommuting ghost fields $c(x)$
with ghost number one, which leads to a nilpotent symmetry operator,
the so-called BRST operator $s$.
The search of BRST invariant Lagrangians and possible anomalies is then
reduced to solve the following cohomology problem
\be
\label{CE}
sa=0~~~,~~~a\neq s\widehat{a} \ ,
\ee
where $s$ is the nilpotent BRST operator and $a$ is the
breaking term of eq.\equ{BREAK}.
In particular, the BRST formalism allows now to characterize classical
actions and anomalies as BRST invariant functionals. Especially, an action
is a BRST invariant functional with ghost number zero while an anomaly
corresponds to a BRST invariant functional with ghost number one.
Eq.\equ{CE} is now the Wess-Zumino consistency relation within the BRST
formalism and restricts strongly the possible solutions of $a$.
For further need we are using the concepts of differential forms, where
$d$ denotes the exterior space-time derivative $d=dx^{\mu}\6_{\mu}$
and where $a$ is described by an integrated volume form ($N$-form) in
$N$ space-time dimension
\be
a=\int\AA \ ,
\ee
with $\AA$ local polynomial.
The condition \equ{CE} implies the following local equation
\be
\label{LE}
s\AA+d\QQ=0 \ ,
\ee
where $\QQ$ is some local polynomial.
The exterior space-time derivative $d$ and the BRST operator $s$ fulfill
\be
\label{NIL}
s^{2}=d^{2}=sd+ds=0 \ .
\ee
$\AA$ is said non-trivial if
\be
\AA \not= s\hat\AA+d\hat\QQ \ ,
\ee
with $\hat\AA$ and $\hat\QQ$ local polynomials.
In this case the integral of $\AA$ on space-time, $\int\AA$,
identifies a cohomology class of the BRST operator $s$ and, according
to its ghost number, it corresponds to an invariant Lagrangian
(ghost number zero) or to an anomaly (ghost number one).

The local
equation \equ{LE}, due to the relations \equ{NIL} and to the
familiar algebraic
Poincar\'e Lemma~\cite{cotta,dragon}
\be
\label{POINCARE}
d\O=0~~ \Longleftrightarrow ~~\O=d\widehat{\O}+d^{N}\!x~\LL+const \ ,
\ee
is easily seen to generate a tower of
descent equations
\bea
\label{LADDER}
&&s\QQ+d\QQ^{1}=0 \ , \non
&&s\QQ^{1}+d\QQ^{2}=0 \ , \non
&&s\QQ^{2}+d\QQ^{3}=0 \ , \non
&&~~~~~.....\non
&&~~~~~.....\non
&&s\QQ^{k-1}+d\QQ^{k}=0 \ , \non
&&s\QQ^{k}=0 \ ,
\eea
with $\QQ^{i}$ local field polynomials. The index $i$ describes the grading
of the local polynomials $\QQ^{i}$ (see Section 3).

As it has been well-known for several years, these equations can be solved
by using a transgression procedure based on the so-called
{\it Russian formula}
\cite{witten,baulieu,dviolette,tmieg,bandelloni,ginsparg,tonin1,stora,brandt}.
More recently an alternative way of finding non-trivial solutions of
the ladder \equ{LADDER} has been proposed by S.P. Sorella and has been
successfully applied to the study of the Yang-Mills gauge
anomalies~\cite{silvio}.
The method is based on the introduction of an operator $\d$ which
allows the expression of the exterior derivative $d$ as a BRST commutator,
i.e.:
\be
\label{DECOMP}
d=-[s,\d] \ .
\ee
One easily verifies that, once the decomposition \equ{DECOMP} has been
found, successive applications of the operator $\d$ on the polynomial
$\QQ^{k}$ which solves the last equation of the tower
\equ{LADDER} give an explicit non-trivial solution for the higher
cocycles $\QQ^{k-1}, ....., \QQ^{1},
\QQ,$ and $\AA$.

Actually, the decomposition \equ{DECOMP}
represents one of the most interesting features of the topological
field theories~\cite{schwarz,birmingham} and of the bosonic string
and superstring in the Beltrami and Super-Beltrami
parametrization~\cite{manfred}.
A remarkable fact is also that solving the last equation of the tower
\equ{LADDER}
is a problem of local BRST cohomology instead of a modulo-$d$ one.
One sees then that, due to the operator $\d$, the study of the
cohomology of $s$ modulo $d$ is essentially
reduced to the study of the local cohomology of $s$ which, in turn,
can be systematically analyzed by using the powerful technique of the
spectral sequences~\cite{dixon}. Actually, as
proven in~\cite{tataru}, the solutions
obtained by utilizing the decomposition \equ{DECOMP} turn out to be
completely equivalent to that based on the {\it Russian formula}, i.e.
they differ only by trivial cocycles.

The aim of this work is twofold. In a first step it will be demonstrated that
the decomposition \equ{DECOMP} can be successfully applied to gravity
including local Lorentz rotations, diffeomorphisms, and Weyl transformations,
and that it holds also in the presence of torsion.
In the second step, we will see that the operator $\d$ gives an elegant and
straightforward way of classifying the cohomology classes of the
full BRST operator in any space-time dimension.
In particular, the eq.\equ{DECOMP} will allow for a
cohomological interpretation of the cosmological constant, of Lagrangians
for pure Einstein gravity and generalizations including also torsion.
Additionally, Chern-Simons terms, gravitational and Weyl anomalies
are considered. The last point is devoted to the discussion of the
coupling of a scalar matter field to gravity with and without Weyl symmetry.

This work is a continuation of a previous one~\cite{werneck},
where the decomposition \equ{DECOMP}
was shown to hold in the case of pure Lorentz transformations involving
only the Lorentz connection $\o$ and the Riemann tensor $R$ and without
taking into account the explicit presence of the vielbein $e$ and of
the torsion $T$. More recently the decomposition \equ{DECOMP}
was used to investigate the cohomological problem of gravity with
torsion~\cite{otmar}.

The further analysis is based on the geometrical formalism
introduced by L. Baulieu and J. Thierry-Mieg~\cite{baulieu,tmieg} which
allows to reinterpret the BRST transformations
as Maurer-Cartan horizontality conditions.
In particular, this formalism turns out to be very useful in the case
of gravity~\cite{baulieu,tmieg}, since it naturally includes the torsion.
In addition, it allows to formulate the diffeomorphism transformations
as local translations in the tangent space by means of the introduction
of the ghost field $\h^{a}=\x^{\mu}e_{\mu}^{a}$ where $\x^{\mu}$
denotes the usual diffeomorphism ghost and $e_{\mu}^{a}$ is the
vielbein.

We recall also that the BRST formulation of gravity with torsion
has already been proposed by~\cite{okubo,lee} in order to
study the quantum aspects of gravity.
In particular, the authors of~\cite{lee} discussed a four dimensional
torsion Lagrangian, with $GL(4,R)$ as the gauge group,
which is able to reproduce the Einstein gravity in the low
energy limit. These BRST transformations could be taken as the
starting point for a purely cohomological algebraic analysis without any
reference to a particular Lagrangian.
Furthermore, our choice of adopting the Maurer-Cartan formalism is
due to the fact that when combined with the introduction of the
translation ghost $\h^{a}$ it will give
us the possibility of a fully tangent space formulation of gravity.

This step, as we shall see in details, will allow to
introduce the decomposition \equ{DECOMP} in a very simple way
and will produce an elegant and compact formula (see Section 5)
for expressing the whole solution of the BRST descent equations, our
aim is that of giving a cohomological interpretation of the
gravitational Lagrangians and of the anomalies in any
space-time dimension with and without Weyl symmetry.
Moreover, the explicit presence of the torsion $T$ and of the
translation ghost $\h^{a}$ gives the possibility of
introducing an algebraic BRST setup which turns out to be
different from that obtained from the analysis of Brandt et al.~\cite{brandt},
where similar techniques have been used.

Finally, we stress that the main purposes of the present work
are dedicated on one hand to discuss the solutions of the local
gravitational cohomology problem without Weyl symmetry
\be
sa(e,\o,R,T)=0 \ ,
\ee
and on the other hand to find solutions of the full local
cohomology problem including Weyl symmetry
\be
sa(e,\o,A,R,T,F)=0 \ ,
\ee
where the vielbein field $e$, the Lorentz connection $\o$, the abelian Weyl
gauge
field $A$, the Riemann tensor $R$, the torsion $T$, and the Weyl curvature $F$
are
treated as unquantized classical fields, as done in~\cite{werneck},
which when coupled to some matter fields (scalars or fermions)
give rise to an effective action whose quantum expansion reduces to
the one-loop order.

The work is organized as follows:
In the several parts of Section 2 we
briefly mention some basic elements, respectively in Section 2.1 gravity
with torsion, in Section 2.2 Weyl transformations, in Section 2.3 the
powerful BRST formalism, and finally in Section 2.4 the elegant technique of
differential forms. After this, we will introduce in Section 3 the so-called
Maurer-Cartan horizontality conditions for gravity with torsion, for
Weyl transformations, and for scalar matter fields. In particular, in Section
3.3
the BRST transformations for local Lorentz rotations, diffeomorphisms,
and Weyl transformations are derived in a complete tangent space formalism.
In Section 4 the operator $\d$ is introduced and we show how it can be used to
solve the descent equations \equ{LADDER}.
Section 5 is dedicated to the study of some explicit examples, like the
cosmological constant, the Einstein and the generalized torsion Lagrangians
as well as the Chern-Simons terms, the gravitational and Weyl anomalies,
and the matter field Lagrangians.
Section 6 deals with the geometrical meaning of the
decomposition \equ{DECOMP} and some detailed calculations
can be found in the final appendices, respectively important commutator
relations in the tangent space and the determinant of the vielbein in
connection with the $\ve$ tensor.


\section{Basic elements}


\subsection{Gravitational fields}

It is well-known that the gauge transformations associated with
gravity are general coordinate transformations or diffeomorphisms,
i.e. arbitrary reparametrizations of the $N$-dimensional
space-time, and local
Lorentz rotations in a tangent space of dimension $N$
\footnote{As usual, Latin and Greek indices refer to the
tangent space and to the euclidean space-time. Both the world
indices and the local Lorentz indices take values from $1$ to $N$.}.
A diffeomorphism can be written as
\be
\label{GCT}
x^{\mu} \rightarrow \bar{x}^{\mu}=x^{\mu}+\x^{\mu}(x) \ ,
\ee
where $\x^{\mu}(x)$ is an infinitesimal parameter of the
general coordinate transformation.
A scalar field $\ph(x)$ has the property
\be
\label{SCALAR}
\bar{\ph}(\bar{x})=\ph(x) \ ,
\ee
and can be expanded over a point $x$ as follows:
\be
\label{SCALAREXP}
\ph(\bar{x})=\ph(x+\x(x))=\ph(x)+\x^{\mu}(x)\6_{\mu}\ph(x)
+\OO(\x)^{2} \ .
\ee
{}From eqs.\equ{SCALAR} and \equ{SCALAREXP} one can read off the
transformation of a scalar field under general coordinate
transformations or diffeomorphisms
\be
\label{DIFF-SCALAR}
\d_{D}\ph=\ph(x)-\ph(\bar{x})
=-\x^{\mu}\6_{\mu}\ph=\LL_{\x}\ph \ ,
\ee
where $\LL_{\x}$ denotes the Lie derivative~\cite{sexl}.
A covariant (contravariant) vector field $V_{\mu}$ ($V^{\mu}$)
transforms under diffeomorphisms as
\bea
\label{DIFF-TENSOR}
\d_{D}V_{\mu} \= -\x^{\l}\6_{\l}V_{\mu}
-(\6_{\mu}\x^{\l})V_{\l} = \LL_{\x}V_{\mu} \ , \non
\d_{D}V^{\mu} \= -\x^{\l}\6_{\l}V^{\mu}
+(\6_{\l}\x^{\mu})V^{\l} = \LL_{\x}V^{\mu} \ ,
\eea
and analogous for tensors of higher ranks.

Now we consider local Lorentz rotations in the tangent
space which act on the fields according to
\be
\label{LOR-ROT}
\Ph(x) \rightarrow \Ph(x)+\frac{1}{2}\e^{ab}(x)M_{ab}\Ph(x) \ ,
\ee
where $M_{ab}$ are the generators of the local Lorentz rotation
group $SO(N)$ in $N$ dimensions in the representation appropriate
to the field $\Ph(x)$ and $\e^{ab}$ are the infinitesimal
parameters of the transformation.
The generators $M_{ab}$ and parameters $\e^{ab}$
are antisymmetric in $(ab)$
\be
\label{ANTISYMMETRY}
M_{ab}=-M_{ba}~~~,~~~\e^{ab}=-\e^{ba} \ .
\ee
An antihermitian representation of the generators $M_{ab}$
is given by
\be
\label{REPRESENT}
(M_{ab})^{c}_{~d}=\d^{c}_{a}\h_{bd}-\d^{c}_{b}\h_{ad} \ ,
\ee
which satisfies the commutator algebra
\be
\label{LOR-GENERATOR}
[M_{ab},M_{cd}]=-\h_{ac}M_{bd}+\h_{ad}M_{bc}+\h_{bc}M_{ad}
-\h_{bd}M_{ac} \ .
\ee
With the help of eq.\equ{REPRESENT} follows
from eq.\equ{LOR-ROT} the transformation of a contravariant tangent
space vector field $V^{a}$ under local Lorentz rotations
\bea
\label{TANGVECT}
\d_{L}V^{a}\=\frac{1}{2}\e^{mn}(M_{mn})^{a}_{~b}V^{b} \non
\=\frac{1}{2}\e^{mn}(\d^{a}_{m}\h_{nb}-\d^{a}_{n}\h_{mb})V^{b} \non
\=\e^{a}_{~b}V^{b} \ ,
\eea
and analogous for a covariant vector field $V_{a}$
\be
\d_{L}V_{a}=\e_{a}^{~b}V_{b} \ .
\ee
Remark that a scalar field $\ph$ has no tangent space index and
therefore does not transform under local Lorentz rotations
\be
\d_{L}\ph=0 \ .
\ee

Now one can define a Lorentz covariant derivative $D_{\mu}$ by
introducing a gauge field associated with the local
Lorentz rotations \equ{LOR-ROT} which is called the Lorentz
connection field $\o^{ab}_{~~\mu}$
\be
\label{LORCOVDER}
D_{\mu}\Ph=\6_{\mu}\Ph+\frac{1}{2}\o^{ab}_{~~\mu}M_{ab}\Ph \ ,
\ee
The covariant derivative transforms according
to eq.\equ{LOR-ROT} as
\be
\d_{L}(D_{\mu}\Ph)=\frac{1}{2}\e^{ab}M_{ab}(D_{\mu}\Ph) \ ,
\ee
which determines with eq.\equ{LOR-GENERATOR}
the transformation property of the Lorentz connection field under
local Lorentz rotations
\be
\label{LOR-OMEGA}
\d_{L}\o^{ab}_{~~\mu}=-\6_{\mu}\e^{ab}-\o^{a}_{~c\mu}\e^{cb}
+\o^{b}_{~c\mu}\e^{ca} \ .
\ee
Remark that the Lorentz connection field $\o^{ab}_{~~\mu}$ carries
both world and tangent space indices. The presence of a world
index is characteristic for gauge fields, since these fields
are introduced in order to compensate for the effects caused by
the local gauge invariance transformations.

Up to now the derivative \equ{LORCOVDER} is
covariant under local Lorentz rotations but
not covariant under diffeomorphisms. In other
words the variation of $D_{\mu}\Ph$ does not only depend on the
parameters $\x^{\mu}(x)$ taken at the same space-time point,
but it also depends on their values at neighbouring points.
This is reflected in the presence of a derivative of $\x^{\mu}(x)$
in the transformation law
\be
\label{COVD}
\d(D_{\mu}\Ph)=\frac{1}{2}\e^{ab}M_{ab}(D_{\mu}\Ph)
-\x^{\nu}\6_{\nu}(D_{\mu}\Ph)-(\6_{\mu}\x^{\nu})(D_{\nu}\Ph) \ ,
\ee
with
\be
\d=\d_{D}+\d_{L} \ .
\ee
{}From eqs.\equ{LOR-OMEGA} and \equ{COVD} one gets the transformation
of the Lorentz connection field $\o^{ab}_{~~\mu}$ under diffeomorphisms
\be
\label{DIFF-OMEGA}
\d_{D}\o^{ab}_{~~\mu}=-\x^{\nu}\6_{\nu}\o^{ab}_{~~\mu}
-(\6_{\mu}\x^{\nu})\o^{ab}_{~~\nu} \ ,
\ee
which is the usual transformation law of a covariant vector field
under general coordinate transformations \equ{DIFF-TENSOR}.
To remove $\6_{\mu}\x^{\nu}$ in the transformation rule \equ{COVD}
one introduces another field whose variation can compensate for
the undesired term. Since $D_{\mu}\Ph$ still carries a world index
the new type of field should convert world indices into local
Lorentz indices. This leads to the introduction of the
field $E_{a}^{\mu}$, which can be used to define a fully covariant
derivative $D_{a}\Ph$ in the following way
\be
D_{a}\Ph=E^{\mu}_{a}D_{\mu}\Ph \ .
\ee
This derivative no longer carries a world index and transforms
therefore as a vector field under local Lorentz rotations and as
a scalar field under diffeomorphisms
\bea
\d(D_{a}\Ph)\=-\x^{\mu}\6_{\mu}(D_{a}\Ph)+\e_{a}^{~b}(D_{b}\Ph) \non
\=-\x^{\mu}\6_{\mu}(D_{a}\Ph)
+\frac{1}{2}\e^{mn}(M_{mn})_{a}^{~b}(D_{b}\Ph) \ ,
\eea
which determines the transformation behaviour of $E_{a}^{\mu}$
under diffeomorphisms and local Lorentz rotations
\be
\d E_{a}^{\mu}=\e_{a}^{~b}E_{b}^{\mu}-\x^{\l}\6_{\l}E_{a}^{\mu}
+(\6_{\l}\x^{\mu})E_{a}^{\l} \ .
\ee
The field $E_{a}^{\mu}$ not only may be regarded as the gauge
field of general coordinate transformations but also may be seen
in a geometric context. Namely the $N$ contravariant
vector fields $E_{a}^{\mu}$ specify the basis vectors of the linear
tangent space at each point of a $N$-dimensional Riemannian
space-time manifold. This implies that $E_{a}^{\mu}$ is
non-singular and has an inverse $e^{a}_{\mu}$ defined by
\bea
\label{ORTHO}
e^{a}_{\mu}E^{\mu}_{b}\=\d^{a}_{~b} \ , \non
e^{a}_{\nu}E^{\mu}_{a}\=\d^{~\mu}_{\nu} \ .
\eea
The standard nomenclature is to call $e^{a}_{\mu}$ the vielbein
field and $E_{a}^{\mu}$ the inverse vielbein field.
Under diffeomorphisms and local Lorentz rotations the vielbein
field transforms as
\be
\label{TRANS-VIELBEIN}
\d e^{a}_{\mu}=\e^{a}_{~b}e^{b}_{\mu}-\x^{\l}\6_{\l}e^{a}_{\mu}
-(\6_{\mu}\x^{\l})e^{a}_{\l} \ .
\ee
In this context the group of local Lorentz rotations is called
the structure group or tangent space group, which rotates the
local tangent space frame.
The existence of the vielbein field allows the introduction of
a covariant metric tensor on the Riemannian space-time manifold
which is symmetric
\be
\label{BASIC}
g_{\mu\nu}(x)=e^{a}_{\mu}(x)e^{b}_{\nu}(x)\h_{ab}=g_{\nu\mu}(x) \ ,
\ee
where $\h_{ab}$ is a Lorentz invariant tensor, which can be used
for local measurements of distances and angles in space-time.

As usual one extends the concept of covariance under
diffeomorphisms and local Lorentz rotations to quantities that
carry world indices. The construction of a covariant derivative
$\DD_{\mu}$ for world covariant
vector fields $X_{\mu}$ is straightforward. By using the inverse
vielbein field one convert the world indices to tangent space
indices, then
apply the covariant derivative \equ{LORCOVDER}, and
with the help of the vielbein field one reconvert the
Lorentz indices into world indices. Hence one has
\bea
\label{WORLDCOVD}
\DD_{\mu}X_{\nu}\=e^{a}_{\nu}D_{\mu}(E_{a}^{\r}X_{\r}) \ , \non
\DD_{\mu}X^{\nu}\=E_{a}^{\nu}D_{\mu}(e^{a}_{\r}X^{\r}) \ .
\eea
The same argument can be applied to define covariant derivatives
acting directly on world tensors. Eqs.\equ{WORLDCOVD} can be
rewritten in the form
\bea
\label{WORLDCOVD2}
\DD_{\mu}X_{\nu}\=D_{\mu}X_{\nu}-\G_{\mu\nu}^{~~\r}X_{\r} \ , \non
\DD_{\mu}X^{\nu}\=D_{\mu}X^{\nu}+\G_{\mu\r}^{~~\nu}X^{\r} \ ,
\eea
with the so-called affine connection $\G_{\mu\nu}^{~~\r}$
defined by
\be
\label{AFFCON}
\G_{\mu\nu}^{~~\r}=-e^{a}_{\nu}D_{\mu}E_{a}^{\r}
=(D_{\mu}e_{\nu}^{a})E_{a}^{\r} \ .
\ee
Note, that the covariant derivatives in \equ{WORLDCOVD2} and the
affine connection contain the Lorentz connection field
$\o^{ab}_{~~\mu}$. The representation \equ{AFFCON} for the affine
connection satisfies the metric postulate, which asserts that the
metric and thus the vielbein field is covariantly constant.
Indeed it is straightforward to verify that \equ{AFFCON} is
equivalent to
\be
\label{METRIC-CONDITION}
\DD_{\mu}e^{a}_{\nu}=0 \ .
\ee
The affine and Lorentz connection fields are therefore not independent;
furthermore it is easy to show with eq.\equ{BASIC} that
\be
\DD_{\mu}g_{\nu\r}=0 \ ,
\ee
from which one deduces that the affine connection must satisfy
following identity
\be
\label{METRICITY}
\G_{\mu\nu}^{~~\s}g_{\s\r}+\G_{\mu\r}^{~~\s}g_{\nu\s}
=\6_{\mu}g_{\nu\r} \ .
\ee
The so-called curvature or Riemann tensor $R$ and torsion
tensor $T$ are found
by using the Ricci identity~\cite{sexl}, which relates these tensors to
commutators of covariant derivatives
\be
\label{COMMUTATOR}
[D_{m},D_{n}] \equiv \frac{1}{2}R_{~~mn}^{ab}(\o)M_{ab}
-T_{mn}^{a}(e,\o)D_{a} \ .
\ee
Evaluation of the left-hand side of eq.\equ{COMMUTATOR} shows
that the Riemann tensor of the local Lorentz rotations and the
torsion tensor are given by
\bea
\label{RIEMANN}
R^{ab}_{~~\mu\nu}(\o)\=\6_{\mu}\o^{ab}_{~~\nu}
-\6_{\nu}\o^{ab}_{~~\mu}
+\o^{a}_{~c\mu}\o^{cb}_{~~\nu}-\o^{a}_{~c\nu}\o^{cb}_{~~\mu} \ , \\
\label{TORSION}
T^{a}_{\mu\nu}(e,\o)\=D_{\mu}e^{a}_{\nu}-D_{\nu}e^{a}_{\mu} \ .
\eea
According to eq.\equ{METRIC-CONDITION}
and the definition \equ{TORSION} the torsion tensor $T$
corresponds to the antisymmetric part of the affine connection
\be
\label{ANTICONNECTION}
\G_{\mu\nu}^{~~\r}-\G_{\nu\mu}^{~~\r}=E_{a}^{\r}T^{a}_{\mu\nu} \ .
\ee
Alternatively one may also compute the Ricci identity for covariant
derivatives \equ{WORLDCOVD2} applied on a world vector
\be
[\DD_{\mu},\DD_{\nu}]X_{\r} \equiv
\frac{1}{2}R^{ab}_{~~\mu\nu}M_{ab}X_{\r}
-T_{\mu\nu}^{\t}\DD_{\t}X_{\r}-R^{\s}_{~\r\mu\nu}X_{\s} \ .
\ee
With the relation
\be
\label{D-MUNU}
[D_{\mu},D_{\nu}]=\frac{1}{2}R^{ab}_{~~\mu\nu}M_{ab} \ ,
\ee
and eq.\equ{ANTICONNECTION} one obtains the curvature tensor
\be
R_{~\r\mu\nu}^{\s}(\G)=\6_{\mu}\G_{\nu\r}^{~~\s}
-\6_{\nu}\G_{\mu\r}^{~~\s}
+\G_{\mu\r}^{~~\t}\G_{\nu\t}^{~~\s}
-\G_{\nu\r}^{~~\t}\G_{\mu\t}^{~~\s} \ .
\ee
Remark that, however, $R(\G)$ is not an independent object.
By evaluating
\be
\DD_{[\mu}\DD_{\nu]}e^{a}_{\r}=\DD_{\mu}\DD_{\nu}e^{a}_{\r}
-\DD_{\nu}\DD_{\mu}e^{a}_{\r}=0 \ ,
\ee
one verifies that
\be
R_{~\r\mu\nu}^{\s}=R^{a}_{~b\mu\nu}e_{\r}^{b}E_{a}^{\s} \ .
\ee
Taking cyclic permutations of the triple commutator
of covariant derivatives, which satisfies the Jacobi identity
\be
[D_{\mu},[D_{\nu},D_{\r}]]+[D_{\nu},[D_{\r},D_{\mu}]]
+[D_{\r},[D_{\mu},D_{\nu}]]=0 \ ,
\ee
and taking cyclic permutations of the covariant derivative
applied on the definition of the torsion \equ{TORSION}
\be
D_{[\mu}T_{\nu\r]}^{a}=D_{[\mu}D_{\nu}e_{\r]}^{a} \ ,
\ee
yields the well-known Bianchi identities
\bea
\label{BIANCHI-R}
D_{[\mu}R_{~~\nu\r]}^{ab}\=0 \ , \\
\label{BIANCHI-T}
D_{[\mu}T_{\nu\r]}^{a}\=R_{~b[\mu\nu}^{a}e_{\r]}^{b} \ .
\eea
The combination of \equ{METRICITY} and \equ{ANTICONNECTION}
now fully determines the affine connection in terms of the metric
field and the torsion tensor
\be
\label{ACONNECTION}
\G_{\mu\nu}^{~~\r}=\{_{\mu\nu}^{\r}\}
+E_{a}^{\r}e_{\nu b}K_{\mu}^{ab} \ ,
\ee
where $\{_{\mu\nu}^{\r}\}$ denoting the Christoffel symbol which
is symmetric in $(\mu\nu)$
\be
\{_{\mu\nu}^{\r}\}=\frac{1}{2}g^{\r\s}(\6_{\mu}g_{\nu\s}
+\6_{\nu}g_{\mu\s}-\6_{\s}g_{\mu\nu}) \ ,
\ee
and $K^{ab}_{\mu}$ the contorsion tensor which is antisymmetric
in $(ab)$
\be
K^{ab}_{\mu}=\frac{1}{2}E^{a\r}E^{\nu b}(T^{c}_{\mu\nu}e_{\r c}
+T^{c}_{\r\mu}e_{\nu c}-T^{c}_{\nu\r}e_{\mu c}) \ .
\ee
The value of the Lorentz connection field corresponding to
eq.\equ{ACONNECTION} is
\be
\o^{ab}_{~~\mu}=\o^{ab}_{~~\mu}(e)+K^{ab}_{\mu} \ ,
\ee
with
\be
\label{ANHOLONOMY}
\o^{ab}_{~~\mu}(e)=\frac{1}{2}e^{c}_{\mu}(-\O^{ab}_{~~c}
+\O^{b~a}_{~c}+\O^{~ab}_{c}) \ ,
\ee
and $\O^{ab}_{~~c}$ the coefficients of anholonomity which
measures the non-commutativity of the vielbein basis
\be
\O_{ab}^{~~c}=E_{a}^{\mu}E_{b}^{\nu}(\6_{\mu}e_{\nu}^{c}
-\6_{\nu}e_{\mu}^{c}) \ .
\ee
In the absence of torsion the Bianchi identitiy \equ{BIANCHI-T}
takes the simple form
\be
R_{~~[\mu\nu}^{ab}e_{\r]b}=0 \ .
\ee
In this case the Lorentz connection field $\o^{ab}_{~~\mu}$
is given by eq.\equ{ANHOLONOMY}
\be
\o^{ab}_{~~\mu}=\o^{ab}_{~~\mu}(e) \ ,
\ee
and the affine connection $\G_{\mu\nu}^{~~\r}$ is nothing but
the Christoffel symbol
\be
\G_{\mu\nu}^{~~\r}=\{_{\mu\nu}^{\r}\} \ .
\ee
This implies the symmetry of the Riemann tensor under
cyclic permutations of the indices
\be
\label{CYCLICSYMM}
R_{\s\r\mu\nu}+R_{\s\mu\nu\r}+R_{\s\nu\r\mu}=0 \ ,
\ee
and the pair exchange symmetry of the Riemann tensor
\be
R_{\s\r\mu\nu}=R_{\mu\nu\s\r} \ .
\ee
By contracting
the Riemann tensor one can construct two further objects,
the Ricci tensor $R_{~\mu}^{a}$
\be
R^{a}_{~\mu}=R_{~~\mu\nu}^{ab}E_{b}^{\nu} \ ,
\ee
which is in the case of vanishing torsion symmetric
\be
R_{~\mu}^{a}=R_{\mu}^{~a} \ ,
\ee
and the Riemann scalar or curvature scalar $R$
\be
\label{RICCI-SCALAR}
R=R_{~\mu}^{a}E^{\mu}_{a}
=R_{~~\mu\nu}^{ab}E^{\mu}_{a}E^{\nu}_{b} \ .
\ee


\subsection{Weyl transformations}

In this section we will introduce the familiar Weyl transformations
and the connection to diffeomorphisms and local Lorentz rotations.
A local Weyl transformation~\cite{weyl} is a local rescaling of the
metric field $g_{\mu\nu}$ according to
\be
\label{WEYL-TRANS}
g_{\mu\nu}(x) \rightarrow e^{\WW(g_{\mu\nu})\O(x)}g_{\mu\nu}(x) \ ,
\ee
where $\WW(g_{\mu\nu})$ is the so-called Weyl weight of the metric tensor
and $\O(x)$ is the parameter for the Weyl transformation.
As usual, the Weyl weight for the metric tensor $g_{\mu\nu}$
is fixed to the value two
\be
\WW(g_{\mu\nu})=2 \ .
\ee
The infinitesimal transformation is then given by
\be
\label{WEYL-METRIC}
\d_{W}g_{\mu\nu}=2\O g_{\mu\nu} \ ,
\ee
and analogous for an arbitrary field $\Ph$, with an appropriate
Weyl weight, one has
\be
\label{WEYL-FIELD}
\d_{W}\Ph=\WW(\Ph)\O \Ph \ .
\ee
The determinant of the metric field, defined as
\be
g=det(g_{\mu\nu}) \ ,
\ee
has now the Weyl weight
\be
\WW(g)=2N \ ,
\ee
where $N$ denotes the space-time dimension.
{}From eq.\equ{BASIC} one can read off the following identity for the
determinant of the vielbein field $e^{a}_{\mu}$
\be
e=\sqrt{g}~~~,~~~e=det(e^{a}_{\mu}) \ ,
\ee
which implies the Weyl weight for the determinant of the
vielbein field
\be
\WW(e)=N \ .
\ee
With this setup the Weyl weights for the following basic
fields are then fixed to
\bea
\label{WEYL-WEIGHTS}
\WW(e^{a}_{\mu})\=+1  \ , \non
\WW(E_{a}^{\mu})\=-1  \ , \non
\WW(g_{\mu\nu})\=+2  \ , \non
\WW(g^{\mu\nu})\=-2  \ ,
\eea
and the corresponding Weyl transformations are given by
\bea
\label{WEYL-TRANSFORMATIONS}
\d_{W}e^{a}_{\mu}\=\O e^{a}_{\mu}  \ , \non
\d_{W}E_{a}^{\mu}\=-\O E_{a}^{\mu}  \ , \non
\d_{W}g_{\mu\nu}\=2\O g_{\mu\nu}  \ , \non
\d_{W}g^{\mu\nu}\=-2\O g^{\mu\nu}  \ ,
\eea
respectively
\be
\label{WEYL-VIELBEIN}
\d_{W}e=\d_{W}\sqrt{g}=N\O\sqrt{g} \ .
\ee
Remark that the Weyl weights are additive.
The standard procedure to find the Weyl weight for a scalar
field $\ph$ is to assume, that the action of a scalar field
should have Weyl weight zero under a global Weyl transformation.
{}From
\be
\label{SCALAR-ACTION}
\G=\frac{1}{2}\int\!d^{N}\!x\sqrt{g}g^{\mu\nu}
\6_{\mu}\ph\6_{\nu}\ph \ ,
\ee
one gets therefore the Weyl weight for a scalar field
\be
\label{WEYL-WPHI}
\WW(\ph)=-\frac{N-2}{2} \ ,
\ee
and the Weyl transformation of a scalar field
\be
\label{WEYL-PHI}
\d_{W}\ph=-\frac{N-2}{2}\O\ph \ .
\ee
Remark also that the partial derivative $\6_{\mu}$ and the
flat metric $\h_{ab}$ have Weyl weights zero.
The Weyl transformations for the remaining fields are
determined by the definitions of the Riemann tensor \equ{RIEMANN}
and the torsion \equ{TORSION} and are given by
\bea
\label{WEYL-TORSION}
\d_{W}\o^{ab}_{~~\mu}\=0 \ , \non
\d_{W}T^{a}_{\mu\nu}\=\O T^{a}_{\mu\nu} \ , \non
\d_{W}R^{ab}_{~~\mu\nu}\=0 \ .
\eea


However, the action of the scalar field \equ{SCALAR-ACTION} is not generally
Weyl invariant under a {\it local} infinitesimal Weyl transformation.
Only in two space-time dimensions one can find a Weyl invariance:
\bea
\label{SCALAR-ACTION-TRANS}
\d_{W}\G\=\frac{1}{2}\d_{W}\int\!d^{N}\!x\sqrt{g}g^{\mu\nu}
\6_{\mu}\ph\6_{\nu}\ph \non
\=-(\frac{N-2}{2})\int\!d^{N}\!x\sqrt{g}g^{\mu\nu}
(\6_{\mu}\O)\ph\6_{\nu}\ph \ .
\eea
In order to keep the Weyl invariance in arbitrary space-time dimensions
we introduce a gauge field $A_{\mu}$ for the Weyl transformations and the
corresponding Weyl covariant derivative
\be
\label{WEYL-COVD}
\nabla_{\mu}\Ph=\6_{\mu}\Ph+\WW(\Ph) A_{\mu}\Ph \ .
\ee
As usual, the covariant derivative of a field $\Ph$ should
transform as the field $\Ph$ (see eq.\equ{WEYL-FIELD})
\be
\label{WEYL-COVD-NABLA}
\d_{W}(\nabla_{\mu}\Ph)=\WW(\Ph)\O (\nabla_{\mu}\Ph) \ ,
\ee
which implies the proper transformation of the Weyl gauge field
\be
\label{WEYL-GAUGEFIELD}
\d_{W}A_{\mu}=-\6_{\mu}\O \ .
\ee
The correct Weyl invariant scalar field action can be found by
replacing the partial derivative $\6_{\mu}$ in \equ{SCALAR-ACTION}
by the Weyl covariant derivative \equ{WEYL-COVD}:
\be
\label{SCALAR-ACTION-COVD}
\G=\frac{1}{2}\int\!d^{N}\!x\sqrt{g}g^{\mu\nu}
\nabla_{\mu}\ph\nabla_{\nu}\ph \ .
\ee
Now one can easily proof that the action \equ{SCALAR-ACTION-COVD} is
Weyl invariant in any space-time dimension.
{}From \equ{SCALAR-ACTION-COVD} follows
\bea
\label{SCALAR-ACTION-COVD-TRANS}
\d_{W}\G\=\frac{1}{2}\d_{W}\int\!d^{N}\!x\sqrt{g}g^{\mu\nu}
\nabla_{\mu}\ph\nabla_{\nu}\ph \non
\=\frac{1}{2}\int\!d^{N}\!x[(\d_{W}\sqrt{g}g^{\mu\nu})
\nabla_{\mu}\ph\nabla_{\nu}\ph
+2\sqrt{g}g^{\mu\nu}(\d_{W}\nabla_{\mu}\ph)\nabla_{\nu}\ph] \ ,
\eea
which leads with the help of \equ{WEYL-TRANSFORMATIONS},\equ{WEYL-VIELBEIN}
and \equ{WEYL-COVD-NABLA} to the invariance of the action:
\be
\d_{W}\G=0 \ .
\ee

We remark that in the case of gravity with vanishing torsion exists a second
possibility to construct a Weyl invariant scalar field action without using
the Weyl gauge field~\cite{birrell}:
\be
\label{DAVIES}
\G=\frac{1}{2}\int\!d^{N}\!x\sqrt{g}[g^{\mu\nu}
\6_{\mu}\ph\6_{\nu}\ph+\a R\ph^{2}] \ ,
\ee
with the coupling constant
\be
\a=\frac{1}{4}\frac{N-2}{N-1} \ .
\ee
The Weyl transformation of the Riemann scalar $R$ for vanishing torsion
is given by
\be
\label{WEYL-R1}
\d_{W}R=-2\O R + 2(N-1)\Box\O \ .
\ee
Now it is straightforward to check that \equ{DAVIES} is Weyl invariant.
But in the presence of torsion the Weyl transformation of the Riemann scalar
$R$
is changed according to
\be
\label{WEYL-R2}
\d_{W}R=-2\O R \ ,
\ee
and the action \equ{DAVIES} is no more Weyl invariant!


The commutator of two covariant derivatives is related to the Weyl
field strength
\be
[\nabla_{\mu},\nabla_{\nu}]\Ph=\WW(\Ph) F_{\mu\nu}\Ph \ .
\ee
The corresponding Weyl curvature
$F_{\mu\nu}$ is then given by
\be
\label{WEYL-CURVATURE}
F_{\mu\nu}=\6_{\mu}A_{\nu}-\6_{\nu}A_{\mu} \ ,
\ee
which is invariant under Weyl transformations
\be
\d_{W}F_{\mu\nu}=0 \ ,
\ee
and obeys the Bianchi identity for the Weyl curvature
\be
\6_{[\mu}F_{\nu\r]}=0 \ .
\ee

In order to incorporate also Weyl transformations with diffeomorphisms
and local Lorentz rotations one introduces the following covariant derivative
\be
\label{GENERAL-COVD}
{\bf{D}}_{\mu}=\6_{\mu}+\frac{1}{2}\o^{ab}_{~~\mu}M_{ab}
+\WW A_{\mu} \ .
\ee
The composition of the symmetry transformations, i.e.
diffeomorphisms, local Lorentz rotations, and Weyl
transformations, is done by
\be
\d=\d_{D}+\d_{L}+\d_{W} \ ,
\ee
which leads to the following set of transformations for the basic fields
\bea
\label{SYMTRANS}
\d \ph\=-\x^{\l}\6_{\l}\ph-\frac{N-2}{2}\O\ph \ , \non
\d e^{a}_{\mu}\=\e^{a}_{~b}e^{b}_{\mu}-\x^{\l}\6_{\l}e^{a}_{\mu}
-(\6_{\mu}\x^{\l})e^{a}_{\l}+\O e^{a}_{\mu} \ , \non
\d \o^{a}_{~b\mu}\=-\6_{\mu}\e^{a}_{~b}+\e^{a}_{~c}\o^{c}_{~b\mu}
-\e^{c}_{~b}\o^{a}_{~c\mu}-\x^{\l}\6_{\l}\o^{a}_{~b\mu}
-(\6_{\mu}\x^{\l})\o^{a}_{~b\l} \ , \non
\d A_{\mu}\=-\x^{\l}\6_{\l}A_{\mu}-(\6_{\mu}\x^{\l})A_{\l}
-\6_{\mu}\O \ , \non
\d T^{a}_{\mu\nu}\=\e^{a}_{~b}T^{b}_{\mu\nu}
-\x^{\l}\6_{\l}T^{a}_{\mu\nu}-(\6_{\mu}\x^{\l})T^{a}_{\l\nu}
-(\6_{\nu}\x^{\l})T^{a}_{\mu\l}+\O T^{a}_{\mu\nu} \ , \non
\d R^{a}_{~b\mu\nu}\=\e^{a}_{~c}R^{c}_{~b\mu\nu}
-\e^{c}_{~b}R^{a}_{~c\mu\nu}-\x^{\l}\6_{\l}R^{a}_{~b\mu\nu}
-(\6_{\mu}\x^{\l})R^{a}_{~b\l\nu}
-(\6_{\nu}\x^{\l})R^{a}_{~b\mu\l} \ , \non
\d F_{\mu\nu}\=-\x^{\l}\6_{\l}F_{\mu\nu}
-(\6_{\mu}\x^{\l})F_{\l\nu}-(\6_{\nu}\x^{\l})F_{\mu\l} \ .
\eea
The commutators of the covariant derivatives \equ{COMMUTATOR} and
\equ{D-MUNU} can be generalized to
\bea
\label{D-WEYL}
[{\bf{D}}_{\mu},{\bf{D}}_{\nu}]\Ph=\frac{1}{2}R^{ab}_{~~\mu\nu}M_{ab}\Ph
+\WW(\Ph) F_{\mu\nu}\Ph \ , \non
\eea
and
\be
\label{COMM-WEYL}
[{\bf{D}}_{m},{\bf{D}}_{n}]\Ph=\frac{1}{2}R_{~~mn}^{ab}M_{ab}\Ph
+\WW(\Ph) F_{mn}\Ph-T_{mn}^{a}{\bf{D}}_{a}\Ph \ ,
\ee
where the Riemann tensor $R^{ab}_{~~\mu\nu}$, given by
eq.\equ{RIEMANN}, and the Weyl curvature $F_{\mu\nu}$,
given by eq.\equ{WEYL-CURVATURE}, are unchanged, but the torsion
tensor field $T^{a}_{\mu\nu}$ is now modified, due to the
including of Weyl transformations, according to
\bea
\label{WEYL-TOR}
T^{a}_{\mu\nu}\={\bf{D}}_{\mu}e^{a}_{\nu}
-{\bf{D}}_{\nu}e^{a}_{\mu} \ , \non
\=\nabla_{\mu}e^{a}_{\nu}-\nabla_{\nu}e^{a}_{\mu}
+\o^{a}_{~b\mu}e^{b}_{\nu}-\o^{a}_{~b\nu}e^{b}_{\mu} \ , \non
\=\6_{\mu}e^{a}_{\nu}-\6_{\nu}e^{a}_{\mu}
+\o^{a}_{~b\mu}e^{b}_{\nu}-\o^{a}_{~b\nu}e^{b}_{\mu}
+A_{\mu}e^{a}_{\nu}-A_{\nu}e^{a}_{\mu} \ .
\eea


\subsection{BRST transformations}

The BRST formalism is an elegant and powerful tool for
the consistent describtion of
gauge symmetries in quantum field theory. The standard
procedure is to substitute the infinitesimal parameters of
the several symmetry transformations eqs.\equ{GCT},
\equ{LOR-ROT},\equ{WEYL-TRANS} for the corresponding anticommuting
Faddeev-Popov ghosts
\bea
\x^{\mu} &\rightarrow& \x^{\mu} \ , \non
\e^{ab} &\rightarrow& \th^{ab} \ , \non
\O &\rightarrow& \s \ ,
\eea
where $\x^{\mu}$, $\th^{ab}$, and $\s$ denoting the diffeomorphism
ghost\footnote{Both, for the parameter of diffeomorphisms and
for the diffeomorphism ghost, we use the same symbol.},
the Lorentz ghost, and the Weyl ghost.
{}From the antisymmetry of the parameter $\e^{ab}$ for the local
Lorentz rotations it follows immediately that also the Lorentz
ghost is antisymmetric
\be
\th^{ab}=-\th^{ba} \ .
\ee
All above ghosts have ghost number one. Further the several
symmetry operations are then expressed by a nilpotent anticommuting
operator $s$ which is called the BRST operator
\be
s=s_{D}+s_{L}+s_{W} \ .
\ee
The BRST operator increase the ghost number by one.
The BRST transformations of the
ghosts are constructed in a way that all transformations are
nilpotent. For all the basic fields mentioned so far one has
now the following BRST transformations
\bea
\label{BRST-TRANS}
s\ph\=-\x^{\l}\6_{\l}\ph-\frac{N-2}{2}\s\ph \ , \non
se^{a}_{\mu}\=\th^{a}_{~b}e^{b}_{\mu}-\x^{\l}\6_{\l}e^{a}_{\mu}
-(\6_{\mu}\x^{\l})e^{a}_{\l}+\s e^{a}_{\mu} \ , \non
sE_{a}^{\mu}\=-\th^{b}_{~a}E_{b}^{\mu}-\x^{\l}\6_{\l}E_{a}^{\mu}
+(\6_{\l}\x^{\mu})E_{a}^{\l}-\s E_{a}^{\mu} \ , \non
s\o^{a}_{~b\mu}\=-\6_{\mu}\th^{a}_{~b}+\th^{a}_{~c}\o^{c}_{~b\mu}
-\th^{c}_{~b}\o^{a}_{~c\mu}-\x^{\l}\6_{\l}\o^{a}_{~b\mu}
-(\6_{\mu}\x^{\l})\o^{a}_{~b\l} \ , \non
sA_{\mu}\=-\x^{\l}\6_{\l}A_{\mu}-(\6_{\mu}\x^{\l})A_{\l}
-\6_{\mu}\s \ , \non
sT^{a}_{\mu\nu}\=\th^{a}_{~b}T^{b}_{\mu\nu}
-\x^{\l}\6_{\l}T^{a}_{\mu\nu}-(\6_{\mu}\x^{\l})T^{a}_{\l\nu}
-(\6_{\nu}\x^{\l})T^{a}_{\mu\l}+\s T^{a}_{\mu\nu} \ , \non
sR^{a}_{~b\mu\nu}\=\th^{a}_{~c}R^{c}_{~b\mu\nu}
-\th^{c}_{~b}R^{a}_{~c\mu\nu}-\x^{\l}\6_{\l}R^{a}_{~b\mu\nu}
-(\6_{\mu}\x^{\l})R^{a}_{~b\l\nu}
-(\6_{\nu}\x^{\l})R^{a}_{~b\mu\l} \ , \non
sF_{\mu\nu}\=-\x^{\l}\6_{\l}F_{\mu\nu}
-(\6_{\mu}\x^{\l})F_{\l\nu}-(\6_{\nu}\x^{\l})F_{\mu\l} \ ,
\eea
and
\bea
\label{BRST-GHOST}
s\x^{\mu}\=-\x^{\l}\6_{\l}\x^{\mu} \ , \non
s\th^{a}_{~b}\=\th^{a}_{~c}\th^{c}_{~b}
-\x^{\l}\6_{\l}\th^{a}_{~b} \ , \non
s\s\=-\x^{\l}\6_{\l}\s \ ,
\eea
implying the nilpotency of the BRST operator
\be
s^{2}=0 \ .
\ee
Furthermore the BRST operator commutes with the partial derivative
\be
[s,\6_{\mu}]=0 \ .
\ee

More generally, the consistent treatment of any gauge field model
demands to fix the gauge, in order to guarantee the existence
of the corresponding gauge field propagator. This is achieved
with a Lagrange multiplier field $B$, which forms together with
an antighost field $\bar{c}$ a so-called BRST-doublet
\bea
\label{ANTIGHOST}
s\bar{c}=B \ , \non
sB=0 \ .
\eea
The dependence of this BRST-doublet within the cohomological
problem \equ{CE}
is managed by a useful theorem~\cite{brandt} showing its triviality.


\subsection{Differential forms}

For the sake of clarity and completeness,
this section is devoted to give a brief review of
some properties and definitions of
the well-known and useful formalism of differential forms.
Differential forms simply provide an exceedingly compact
notation for vectors and tensors on an arbitrary manifold.

A scalar function $f(x)$ is called a zero form. One defines
the differential of the zero form $f$ as the one form
\be
df \equiv \frac{\6 f}{\6 x^{\mu}}dx^{\mu} \ ,
\ee
where in $N$ dimensions the index $\mu$ runs from $1$ to $N$.
Thus, the exterior derivative $d$ can be written as
\be
\label{EXTER_DERIV}
d \equiv dx^{\mu}\6_{\mu} \ ,
\ee
which increases the form degree by one and which is a nilpotent operator
\be
d^{2}=0 \ .
\ee
The nilpotency of $d$ is automatically guaranteed
because of the vanishing commutator of two partial derivatives
\be
[\6_{\mu},\6_{\nu}]=0 \ .
\ee

Given a vector function $\Ph_{\mu}$ one constructs the one form
$\Ph$ as follows:
\be
\label{1F}
\Ph \equiv \Ph_{\mu}dx^{\mu} \ .
\ee
The exterior derivative $d$ of the one form \equ{1F} is defined as
\be
d\Ph=\frac{1}{2}(\6_{\mu}\Ph_{\nu}-\6_{\nu}\Ph_{\mu})dx^{\mu}\wedge dx^{\nu} \
,
\ee
where the so-called wedge product is given by
\be
dx^{\mu}\wedge dx^{\nu}=-dx^{\nu}\wedge dx^{\mu} \ .
\ee
Therefore, $d\Ph$ gives the curl of $\Ph$.

In general, given a completely antisymmetric tensor with $p$ indices
$\o_{\mu_{1}\mu_{2}...\mu_{p}}$ one defines a $p$-form as
\be
\o=\frac{1}{p!}\o_{\mu_{1}\mu_{2}...\mu_{p}}
dx^{\mu_{1}}\wedge dx^{\mu_{2}}\wedge ...\wedge dx^{\mu_{p}} \ .
\ee
Clearly, in $N$ dimensions, one cannot have $p$-forms with $p \ge N$
which do not vanish identically.
In order to simplify the notation, one omits the wedge product symbol
and one simply regards $dx^{\mu}$ as an anticommuting Grassmann object.

To illustrate the use of forms, we look at the Yang-Mills theory,
where the gauge field is the one form
\be
A=A_{\mu}dx^{\mu} \ .
\ee
Remark, that here $A_{\mu}=A_{\mu}^{a}\l^{a}$ with generators $\l^{a}$,
and so $A$ is at the same time a form and a matrix.
The curvature associated with the one form $A$ is a two form:
\be
F=dA+AA \ .
\ee
Then the Bianchi identity, expressed in terms of forms, is nothing
but the nilpotency of $d$
\be
dF=(dA)A-A(dA) \ ,
\ee
\be
[A,F]=A(dA)-(dA)A \ .
\ee
With the definition of the covariant derivative one has
\be
DF=dF+[A,F]=0 \ .
\ee

Now we can reformulate the results of the previous sections
in the calculus of differential forms. The corresponding basic
fields are given by the one forms $(e^{a},\o^{a}_{~b},A)$,
$e^{a}$, $\o^{a}_{~b}$, and $A$ being respectively
the vielbein, the Lorentz connection, and the Weyl gauge field
\bea
\label{ONE-FORMS}
e^{a}\=e^{a}_{\mu}dx^{\mu} \ , \non
\o^{a}_{~b}\=\o^{a}_{~b\mu}dx^{\mu}=\o^{a}_{~bm}e^{m} \ , \non
A\=A_{\mu}dx^{\mu}=A_{m}e^{m} \ ,
\eea
and the two forms $(T^{a},R^{a}_{~b},F)$, $T^{a}$, $R^{a}_{~b}$,
and $F$ denoting the torsion, the Riemann tensor, and the
Weyl curvature
\bea
\label{TWO-FORMS}
T^{a}\=\frac{1}{2}T^{a}_{\mu\nu}dx^{\mu}dx^{\nu}
=de^{a}+\o^{a}_{~b}e^{b}+Ae^{a}
=De^{a} \ , \non
R^{a}_{~b}\=\frac{1}{2}R^{a}_{~b\mu\nu}dx^{\mu}dx^{\nu}
=d\o^{a}_{~b}+\o^{a}_{~c}\o^{c}_{~b} \ , \non
F\=\frac{1}{2}F_{\mu\nu}dx^{\mu}dx^{\nu}=dA \ ,
\eea
where
\be
D=d+\o+A
\ee
is the covariant derivative \equ{GENERAL-COVD}.
The remaining zero form fields are the scalar field $\ph$,
the ghost field for diffeomorphisms $\x^{\mu}$, the Lorentz
ghost field $\th^{a}_{~b}$, and the Weyl ghost field $\s$.
{}From eq.\equ{TWO-FORMS} one easily obtains the Bianchi identities
\bea
\label{BI}
DT^{a}\=dT^{a}+\o^{a}_{~b}T^{b}+AT^{a}
=R^{a}_{~b}e^{b}+Fe^{a} , \non
DR^{a}_{~b}\=dR^{a}_{~b}+\o^{a}_{~c}R^{c}_{~b}
-\o^{c}_{~b}R^{a}_{~c}=0 \ , \non
DF\=dF=0 \ .
\eea
Furthermore, one has the anticommutator relation
\be
\label{ANTICOMMUTATORREL}
\{s,d\}=sd+ds=0 \ .
\ee


\section{Maurer-Cartan horizontality conditions}

The aim of this section is to derive the given set of BRST
transformations, defined by the eqs.\equ{BRST-TRANS}-\equ{BRST-GHOST},
from Maurer-Cartan horizontality conditions~\cite{baulieu,tmieg}.
In a first step this geometrical formalism is used to discuss
the simpler case of non-abelian Yang-Mills theory~\cite{baulieu2}.


\subsection{Yang-Mills case}

The BRST transformations of the one form gauge connection
$A^{a}=A^{a}_{\mu}dx^{\mu}$ and the zero form ghost field $c^{a}$
are given by
\bea
\label{BRSTYM}
sA^{a}\=dc^{a}+f^{abc}c^{b}A^{c} \ , \non
sc^{a}\=\frac{1}{2}f^{abc}c^{b}c^{c} \ ,
\eea
with
\be
s^{2}=0 \ ,
\ee
where $f^{abc}$ are the structure constants of the corresponding
gauge group\footnote{Notice that here the indices $a,b,c,...$ are
denoting the gauge group indices.}.
As usual, the adopted grading is given by the sum of the
form degree and of the ghost number. In this sense, the fields $A^{a}$
and $c^{a}$ are both of degree one,
their ghost number being respectively zero and one.
A $p$-form with ghost number $q$ will be denoted by
$\O^{q}_{p}$, its total grading being $(p+q)$.
The two form field strength $F^{a}$ is given by
\be
F^{a}=\frac{1}{2}F^{a}_{\mu\nu}dx^{\mu}dx^{\nu}
=dA^{a}+\frac{1}{2}f^{abc}A^{b}A^{c} \ ,
\ee
and
\be
dF^{a}=f^{abc}F^{b}A^{c} \ ,
\ee
is its Bianchi identity.
In order to reinterpret the BRST transformations \equ{BRSTYM} as a
Maurer-Cartan horizontality condition we introduce the combined
gauge-ghost field
\be
\wti{A}^{a}=A^{a}+c^{a} \ ,
\ee
and the generalized nilpotent differential operator
\be
\label{GDO}
\wti{d}=d-s \ , ~~~\wti{d}^{2}=0 \ .
\ee
Notice that both $\wti{A}^{a}$ and $\wti{d}$ have degree one.
The nilpotency of $\wti{d}$ in \equ{GDO} just implies the nilpotency
of $s$ and $d$, and furthermore fulfills the anticommutator
relation \equ{ANTICOMMUTATORREL}.

Let us introduce also the degree-two field strenght $\wti{F}^{a}$:
\be
\wti{F}^{a}=\wti{d}\wti{A}^{a}
+\frac{1}{2}f^{abc}\wti{A}^{b}\wti{A}^{c} \ ,
\ee
which, from eq.\equ{GDO}, obeys the generalized Bianchi identity
\be
\wti{d}\wti{F}^{a}=f^{abc}\wti{F}^{b}\wti{A}^{c} \ .
\ee
The Maurer-Cartan horizontality condition
reads then
\be
\label{MCYM}
\wti{F}^{a}=F^{a} \ .
\ee
Now it is very easy to check that the BRST transformations
\equ{BRSTYM} can be obtained from the horizontality condition
\equ{MCYM} by simply expanding $\wti{F}^{a}$ in terms of the
elementary fields $A^{a}$ and $c^{a}$ and collecting the terms
with the same form degree and ghost number.
In addition, we remark also the equality leading to the
generalized Bianchi identity
\be
\wti{d}\wti{F}^{a}-f^{abc}\wti{F}^{b}\wti{A}^{c}
=dF^{a}-f^{abc}F^{b}A^{c}=0 \ .
\ee


\subsection{Gravitational case with Weyl symmetry}

To write down the gravitational Maurer-Cartan horizontality conditions
for the model described in the previous section
one introduces a further ghost, as done in~\cite{baulieu,tmieg},
the local translation ghost $\h^{a}$ having ghost number one
and tangent space indices.
As explained in~\cite{tmieg}, the field $\h^{a}$ represents the
ghost of local translations in the tangent space. See also the
discussion of~\cite{mielke} based on an affine approach to gravity.

The local translation ghost $\h^{a}$ can be related~\cite{tmieg} to the
ghost of local diffeomorphism $\x^{\mu}$ by the relation
\be
\label{ETA}
\x^{\mu} = E^{\mu}_{a} \h^{a}~~~,~~~
\h^{a} = \x^{\mu}e^{a}_{\mu} \ ,
\ee
where $E^{\mu}_{a}$ denotes the inverse of the vielbein
$e^{a}_{\mu}$, i.e.
\bea
e^{a}_{\mu}E^{\mu}_{b} \= \d^{a}_{b} \ ,\non
e^{a}_{\mu}E^{\nu}_{a} \= \d^{\nu}_{\mu} \ .
\eea
Proceeding now as for the Yang-Mills case, one defines the nilpotent
differential operator $\wti{d}$ of degree one:
\be
\label{EXTD}
\wti{d}=d-s \ ,
\ee
and the generalized vielbein-ghost field $\ti{e}^{a}$, the extended
Lorentz connection $\wti{\o}^{a}_{~b}$, and the generalized Weyl
gauge field $\wti{A}$
\bea
\label{EVIEL}
\ti{e}^{a} \= e^{a}+\h^{a} \ , \non
\wti{\o}^{a}_{~b} \= \widehat{\o}^{a}_{~b}+\th^{a}_{~b} \ , \non
\wti{A} \= \widehat{A}+\s \ ,
\eea
where $\widehat{\o}^{a}_{~b}$
and $\widehat{A}$ are given by
\bea
\label{HCO}
\widehat{\o}^{a}_{~b}\=\o^{a}_{~bm}\ti{e}^{m}
=\o^{a}_{~b}+\o^{a}_{~bm}\h^{m} \ , \non
\widehat{A}\=A_{m}\ti{e}^{m}
=A+A_{m}\h^{m} \ ,
\eea
with the zero forms $\o^{a}_{~bm}$\footnote{Remark that the zero
form $\o^{a}_{~bm}$ does not possess any symmetric or
antisymmetric property with respect to the lower indices $(bm)$.}
and $A_{m}$ defined by the expansion of the zero form
connection $\o^{a}_{~b\mu}$ and the zero form Weyl gauge field
$A_{\m}$ in terms of the vielbein $e^{a}_{\mu}$,
i.e.:
\bea
\label{SPINC}
\o^{a}_{~b\mu}\=\o^{a}_{~bm}e^{m}_{\mu} \ , \non
A_{\mu}\=A_{m}e^{m}_{\mu} \ .
\eea
As it is well-known, the last formulas stem from the fact that the
vielbein formalism allows to transform locally the space-time
indices of an arbitrary
tensor $\NN_{\mu\nu\r\s...}$ into flat tangent space indices
$\NN_{abcd...}$ by means of the expansion
\be
\label{WORLD}
\NN_{\mu\nu\r\s...}=\NN_{abcd...}
e^{a}_{\mu}e^{b}_{\nu}e^{c}_{\r}e^{d}_{\s}...  \ .
\ee
Vice versa one has
\be
\label{TANG}
\NN_{abcd...}=\NN_{\mu\nu\r\s...}
E^{\mu}_{a}E^{\nu}_{b}E^{\r}_{c}E^{\s}_{d}...  \ .
\ee
According to eqs.\equ{TWO-FORMS}, the generalized torsion field,
the generalized Riemann tensor, and the generalized Weyl curvature
are given by
\bea
\label{DEFTWO}
\wti{T}^{a} \= \wti{d}\ti{e}^{a}+\wti{\o}^{a}_{~b}\ti{e}^{b}+\wti{A}\ti{e}^{a}
=\wti{D}\ti{e}^{a} \ , \non
\wti{R}^{a}_{~b} \= \wti{d}\wti{\o}^{a}_{~b}+\wti{\o}^{a}_{~c}
\wti{\o}^{c}_{~b} \ , \non
\wti{F}\=\wti{d}\wti{A} \ ,
\eea
and are easily seen to obey the generalized Bianchi identities
\bea
\label{GBI}
\wti{D}\wti{T}^{a} \= \wti{d}\wti{T}^{a}
+\wti{\o}^{a}_{~b}\wti{T}^{b}+\wti{A}\wti{T}^{a}
=\wti{R}^{a}_{~b}\ti{e}^{b}+\wti{F}\ti{e}^{a} \ , \non
\wti{D}\wti{R}^{a}_{~b} \= \wti{d}\wti{R}^{a}_{~b}
+\wti{\o}^{a}_{~c}\wti{R}^{c}_{~b}
-\wti{\o}^{c}_{~b}\wti{R}^{a}_{~c}=0 \ , \non
\wti{D}\wti{F}\=\wti{d}\wti{F}=0 \ ,
\eea
with
\be
\wti{D}=\wti{d}+\wti{\o}+\wti{A}
\ee
the generalized covariant derivative.

With these definitions the Maurer-Cartan horizontality conditions
for gravity (with non-vanishing torsion) in the presence of a scalar
field may be expressed in the following way:
{\it $\ti{e}$ and all its generalized covariant
exterior differentials can be expanded over $\ti{e}$ with
classical coefficients},
\bea
\label{MCG1}
\ti{e}^{a}\=\d^{a}_{b}\ti{e}^{b} \equiv horizontal \ , \\
\label{MCG2}
\wti{T}^{a}(\ti{e},\wti{\o})
\=\frac{1}{2}T^{a}_{mn}(e,\o)\ti{e}^{m}\ti{e}^{n}
\equiv horizontal \ , \\
\label{MCG3}
\wti{R}^{a}_{~b}(\wti{\o})
\=\frac{1}{2}R^{a}_{~bmn}(\o)\ti{e}^{m}\ti{e}^{n}
\equiv horizontal \ , \\
\label{MCG4}
\wti{F}(\wti{A})
\=\frac{1}{2}F_{mn}(A)\ti{e}^{m}\ti{e}^{n}
\equiv horizontal \ , \\
\label{MCG5}
\wti{D}\ph
\=(D_{m}\ph)\ti{e}^{m}=(\6_{m}\ph-\frac{N-2}{2} A_{m}\ph)\ti{e}^{m}
\equiv horizontal \ .
\eea
Through eq.\equ{WORLD}, the zero forms $T^{a}_{mn}$,
$R^{a}_{~bmn}$, and $F_{mn}$
are defined by the vielbein expansion of the two forms of the
torsion, the Riemann tensor, and the Weyl curvature of
eq.\equ{TWO-FORMS},
\bea
\label{TWOFORM}
T^{a}\=\frac{1}{2}T^{a}_{mn}e^{m}e^{n} \ , \non
R^{a}_{~b}\=\frac{1}{2}R^{a}_{~bmn}e^{m}e^{n} \ , \non
F\=\frac{1}{2}F_{mn}e^{m}e^{n} \ ,
\eea
and the zero form of the covariant derivative $D_{m}$ is given by
(see also later)
\be
D=e^{m}D_{m} \ .
\ee
Notice also that eqs.\equ{HCO} are nothing but the horizontality
conditions for the Lorentz connection and the Weyl gauge field
expressing the fact that
$\widehat{\o}$ and $\widehat{A}$
themselfs can be expanded over $\ti{e}$.

Eqs.\equ{MCG1}-\equ{MCG5} define the Maurer-Cartan horizontality
conditions for the gravitational case in the presence of
scalar fields and, when expanded in terms of
the elementary fields
$(e^{a}, \o^{a}_{~b}, A, \h^{a}, \th^{a}_{~b}, \s)$, give the
nilpotent BRST transformations corresponding to the diffeomorphism
transformations, the local Lorentz rotations, and
the Weyl transformations.

For a better understanding of this point let us first discuss in
detail the horizontality conditions \equ{MCG5} for the scalar
field and \equ{MCG2} for the torsion.
Making use of eqs.\equ{EXTD}, \equ{EVIEL}, \equ{HCO} and of the
definition \equ{DEFTWO}, one verifies that eq.\equ{MCG5} gives
\be
d\ph-s\ph-\frac{N-2}{2} A\ph-\frac{N-2}{2} A_{m}\h^{m}\ph-\frac{N-2}{2}\s\ph
=(D_{m}\ph)e^{m}+(D_{m}\ph)\h^{m} \ ,
\ee
and eq.\equ{MCG2} leads to
\bea
&&de^{a}-se^{a}+d\h^{a}-s\h^{a}+\o^{a}_{~b}e^{b}+\th^{a}_{~b}e^{b}\non
&&+~\o^{a}_{~b}\h^{b}+\th^{a}_{~b}\h^{b}
+\o^{a}_{~bm}\h^{m}e^{b}+\o^{a}_{~bm}\h^{m}\h^{b}\non
&&+~Ae^{a}+A_{m}\h^{m}e^{a}+\s e^{a}+A\h^{a}+A_{m}\h^{m}\h^{a}
+\s\h^{a}\non
&&=\frac{1}{2}T^{a}_{mn}e^{m}e^{n}
+T^{a}_{mn}e^{m}\h^{n}+\frac{1}{2}T^{a}_{mn}\h^{m}\h^{n} \ ,
\eea
from which, collecting the terms with the same form degree and
ghost number, one easily obtains the BRST transformations for the
scalar field $\ph$, the vielbein $e^{a}$, and for the ghost $\h^{a}$:
\bea
\label{BRSTE}
s\ph\=-\h^{m}\6_{m}\ph-\frac{N-2}{2}\s\ph \ , \non
se^{a}\=d\h^{a}+\o^{a}_{~b}\h^{b}+\th^{a}_{~b}e^{b}
+\o^{a}_{~bm}\h^{m}e^{b}+A_{m}\h^{m}e^{a}+\s e^{a} \non
\+A\h^{a}-T^{a}_{mn}e^{m}\h^{n} \ ,\non
s\h^{a}\=\th^{a}_{~b}\h^{b}+\o^{a}_{~bm}\h^{m}\h^{b}
+A_{m}\h^{m}\h^{a}+\s\h^{a}
-\frac{1}{2}T^{a}_{mn}\h^{m}\h^{n} \ .
\eea
These equations, when rewritten in terms of the variable $\x^{\mu}$ of
eq.\equ{ETA}, take the more familiar forms (see eqs.\equ{BRST-TRANS})
\bea
s\ph\=-\x^{\l}\6_{\l}\ph-\frac{N-2}{2}\s\ph \ , \non
se^{a}_{\mu}\=\th^{a}_{~b}e^{b}_{\mu}+\LL_{\x}e^{a}_{\mu}
+\s e^{a}_{\mu} \ ,\non
s\x^{\mu}\=-\x^{\l}\6_{\l}\x^{\mu} \ ,
\eea
where $\LL_{\x}$ denotes the ordinary Lie derivative along the
direction $\x^{\mu}$, i.e.
\be
\LL_{\x}e^{a}_{\mu}=-\x^{\l}\6_{\l}e^{a}_{\mu}
-(\6_{\mu}\x^{\l})e^{a}_{\l} \ .
\ee
It is now apparent that eq.\equ{BRSTE} represents the tangent space
formulation of the usual BRST transformations corresponding to local
Lorentz rotations, diffeomorphisms, and Weyl transformations
for the scalar field $\ph$, the one form vielbein field $e^{a}$,
and the zero form translation ghost field $\h^{a}$.

One sees then that the Maurer-Cartan horizontality conditions
\equ{MCG1}-\equ{MCG5} together with eq.\equ{DEFTWO} carry in a very
simple and compact form all the informations
concerning the symmetry content with respect to the BRST formalism.
It is quite easy indeed to expand
eqs.\equ{MCG1}-\equ{MCG5} in terms of $e^{a}$ and $\h^{a}$ and work
out the BRST transformations of the remaining fields
$(\o^{a}_{~b}, A, R^{a}_{~b}, T^{a}, F, ...)$.

However, in view of the fact that we will use as fundamental variables
the zero forms $(\o^{a}_{~bm}, A_{m}, R^{a}_{~bmn}, T^{a}_{mn}, F_{mn})$
and the one form $e^{a}$
rather than the one form Lorentz connection $\o^{a}_{~b}$, the one form
Weyl gauge field $A$, and the two forms
$R^{a}_{~b}$, $T^{a}$, and $F$, let us proceed by introducing the partial
derivative $\6_{a}$ of the tangent space.
According to the formulas \equ{WORLD} and \equ{TANG}, the latter is
defined by
\be
\6_{a} \equiv E^{\mu}_{a}\6_{\mu} \ ,
\ee
and
\be
\6_{\mu} = e^{a}_{\mu}\6_{a} \ ,
\ee
so that the intrinsic exterior differential $d$ becomes
\be
d=dx^{\mu}\6_{\mu}=e^{a}\6_{a} \ ,
\ee
and analogous for the covariant derivative $D$
\be
D=dx^{\mu}D_{\mu}=e^{m}D_{m} \ .
\ee
The introduction of the operator $\6_{a}$ and
the use of the zero forms $(\o^{a}_{~bm}, A_{m},$ $R^{a}_{~bmn},$
$T^{a}_{mn}, F_{mn})$ and the one form $e^{a}$
allows for a complete tangent space formulation.
This step, as we shall see later, turns out to be very
useful in the analysis of the corresponding BRST cohomology. Moreover,
as one can easily understand, the knowledge of the BRST transformations
of the zero form sector $(\o^{a}_{~bm}, A_{m}, R^{a}_{~bmn},
T^{a}_{mn}, F_{mn})$
together with the expansions \equ{SPINC}, \equ{TWOFORM} and the
equation \equ{BRSTE} completely characterize the transformation
law of the forms $(\o^{a}_{~b}, A, R^{a}_{~b}, T^{a}, F)$.

For completeness,
now we will discuss the remaining Maurer-Cartan horizontality
conditions and we will derive the BRST transformations of the
corresponding fields. From eq.\equ{MCG4} follows
\bea
&&dA-sA+d(A_{m}\h^{m})-s(A_{m}\h^{m})+d\s-s\s\non
&&=\frac{1}{2}F_{mn}e^{m}e^{n}
+F_{mn}e^{m}\h^{n}+\frac{1}{2}F_{mn}\h^{m}\h^{n} \ ,
\eea
from which, collecting again the terms with the same form degree and
ghost number, one can read off the BRST transformations for the
one form Weyl gauge field $A$ and for the zero form ghost field
$\widehat{\s}$:
\bea
\label{BRSTE2}
sA\=d\widehat{\s}-F_{mn}e^{m}\h^{n} \ , \non
\label{BRSTE3}
s\widehat{\s}\=-\frac{1}{2}F_{mn}\h^{m}\h^{n} \ ,
\eea
where $\widehat{\s}$ is given by the combination
\be
\widehat{\s}=A_{m}\h^{m}+\s \ .
\ee
To determine the BRST transformation for the zero form Weyl gauge
field $A_{m}$ one needs the partial derivative of it in the tangent
space. This can be found by the definition for the two form Weyl
curvature $F$ \equ{TWO-FORMS}
\bea
dA\=F=\frac{1}{2}F_{mn}e^{m}e^{n} \non
\=d(A_{m}e^{m})=\frac{1}{2}(\6_{m}A_{n}-\6_{n}A_{m})e^{m}e^{n}
+A_{m}(de^{m}) \ .
\eea
By inserting the definition of the torsion two form $T^{a}$
\equ{TWO-FORMS} above equation leads to
\be
\label{PARTDERIV-A}
\6_{m}A_{n}-\6_{n}A_{m}=F_{mn}-A_{k}T^{k}_{mn}-A_{k}\o^{k}_{~mn}
+A_{k}\o^{k}_{~nm} \ .
\ee
With the help of eq.\equ{BRSTE} one easily finds
from eq.\equ{BRSTE2} the BRST transformation of the
zero form Weyl gauge field $A_{m}$ according to
\be
\label{BRSTE-A}
sA_{m}=-\h^{k}\6_{k}A_{m}-\th^{k}_{~m}A_{k}-\6_{m}\s-A_{m}\s \ ,
\ee
and by using the eqs.\equ{BRSTE}, \equ{PARTDERIV-A}, and \equ{BRSTE-A}
one gets the BRST transformation of the Weyl ghost $\s$
from eq.\equ{BRSTE3}
\be
\label{BRSTE-SIGMA}
s\s=-\h^{k}\6_{k}\s \ .
\ee
These equations, when rewritten in terms of the variable $\x^{\mu}$,
take the form (see eqs.\equ{BRST-TRANS})
\bea
sA_{\mu}\=-\x^{\l}\6_{\l}A_{\mu}-(\6_{\mu}\x^{\l})A_{\l}
-\6_{\mu}\s \ , \non
s\s\=-\x^{\l}\6_{\l}\s \ .
\eea
Finally, the Maurer-Cartan horizontality condition
for the Riemann tensor \equ{MCG3} gives
\bea
&&d\o^{a}_{~b}-s\o^{a}_{~b}+d(\o^{a}_{~bm}\h^{m})
-s(\o^{a}_{~bm}\h^{m})+d\th^{a}_{~b}-s\th^{a}_{~b} \non
&&+~\o^{a}_{~c}\o^{c}_{~b}+\o^{a}_{~c}\th^{c}_{~b}
+\th^{a}_{~c}\o^{c}_{~b}+\th^{a}_{~c}\th^{c}_{~b} \non
&&+~\o^{a}_{~c}\o^{c}_{~bm}\h^{m}
+\o^{a}_{~cm}\h^{m}\o^{c}_{~b}
+\o^{a}_{~cm}\h^{m}\o^{c}_{~bn}\h^{n} \non
&&+~\th^{a}_{~c}\o^{c}_{~bm}\h^{m}+\o^{a}_{~cm}\h^{m}\th^{c}_{~b} \non
&&=\frac{1}{2}R^{a}_{~bmn}e^{m}e^{n}
+R^{a}_{~bmn}e^{m}\h^{n}+\frac{1}{2}R^{a}_{~bmn}\h^{m}\h^{n} \ ,
\eea
from which the BRST transformations for the
one form Lorentz connection field $\o^{a}_{~b}$ and for the zero
form ghost field $\widehat{\th}^{a}_{~b}$ are found to
\bea
\label{BRSTE4}
s\o^{a}_{~b}\=d\widehat{\th}^{a}_{~b}+\widehat{\th}^{a}_{~c}\o^{c}_{~b}
-\widehat{\th}^{c}_{~b}\o^{a}_{~c}-R^{a}_{~bmn}e^{m}\h^{n} \ , \non
\label{BRSTE5}
s\widehat{\th}^{a}_{~b}\=\widehat{\th}^{a}_{~c}\widehat{\th}^{c}_{~b}
-\frac{1}{2}R^{a}_{~bmn}\h^{m}\h^{n} \ ,
\eea
where $\widehat{\th}^{a}_{~b}$ is defined by the combination
\be
\widehat{\th}^{a}_{~b}=\o^{a}_{~bm}\h^{m}+\th^{a}_{~b} \ .
\ee
The partial derivative of the zero form Lorentz connection $\o^{a}_{~bm}$
follows from the definition of the two form Riemann tensor $R^{a}_{~b}$
\equ{TWO-FORMS}
\bea
d\o^{a}_{~b}\=R^{a}_{~b}-\o^{a}_{~c}\o^{c}_{~b}
=\frac{1}{2}R^{a}_{~bmn}e^{m}e^{n}
-\o^{a}_{~cm}\o^{c}_{~bn}e^{m}e^{n} \non
\=d(\o^{a}_{~bn}e^{n})=\frac{1}{2}(\6_{m}\o^{a}_{~bn}
-\6_{n}\o^{a}_{~bm})e^{m}e^{n}+\o^{a}_{~bm}(de^{m}) \ ,
\eea
which leads to
\bea
\label{PARTDERIV-OMEGA}
\6_{m}\o^{a}_{~bn}-\6_{n}\o^{a}_{~bm}\=R^{a}_{~bmn}
-\o^{a}_{~cm}\o^{c}_{~bn}+\o^{a}_{~cn}\o^{c}_{~bm}
-\o^{a}_{~bk}T^{k}_{mn} \non
\+\o^{a}_{~bk}\o^{k}_{~nm}
-\o^{a}_{~bk}\o^{k}_{~mn}+\o^{a}_{~bn}A_{m}-\o^{a}_{~bm}A_{n} \ .
\eea
The BRST transformations for the zero form Lorentz connection field
$\o^{a}_{~bm}$ and for the Lorentz ghost $\th^{a}_{~b}$ are then
determined by
\bea
\label{BRSTE-OMEGA}
s\o^{a}_{~bm}\=-\h^{k}\6_{k}\o^{a}_{~bm}-\6_{m}\th^{a}_{~b}
+\th^{a}_{~c}\o^{c}_{~bm}-\th^{c}_{~b}\o^{a}_{~cm}
-\th^{k}_{~m}\o^{a}_{~bk}-\s\o^{a}_{~bm} \ , \\
\label{BRSTE-THETA}
s\th^{a}_{~b}\=-\h^{k}\6_{k}\th^{a}_{~b}
+\th^{a}_{~c}\th^{c}_{~b} \ .
\eea
By using the variable $\x^{\mu}$
above BRST transformations correspond to (see eqs.\equ{BRST-TRANS})
\bea
s\o^{a}_{~b\mu}\=-\x^{\l}\6_{\l}\o^{a}_{~b\mu}
-(\6_{\mu}\x^{\l})\o^{a}_{~b\l}-\6_{\mu}\th^{a}_{~b}
+\th^{a}_{~c}\o^{c}_{~b\mu}-\th^{c}_{~b}\o^{a}_{~c\mu} \ , \non
s\th^{a}_{~b}\=-\x^{\l}\6_{\l}\th^{a}_{~b}
+\th^{a}_{~c}\th^{c}_{~b} \ .
\eea
The BRST transformations for the two forms $(T^{a}, R^{a}_{~b}, F)$,
the torsion, the Riemann tensor, and the Weyl curvature, can be
worked out from the generalized Bianchi identities \equ{GBI}
\bea
\label{BRST-TWOFORM-T}
sT^{a}\=(dT^{a}_{mn})e^{m}\h^{n}-T^{a}_{mn}e^{m}(d\h^{n})
+\th^{a}_{~b}T^{b}+\s T^{a}+T^{a}_{mn}T^{m}\h^{n} \non
\-T^{a}_{mn}\o^{m}_{~k}e^{k}\h^{n}+\o^{a}_{~bk}\h^{k}T^{b}
+\o^{a}_{~b}T^{b}_{mn}e^{m}\h^{n}+A_{k}\h^{k}T^{a} \non
\-R^{a}_{~b}\h^{b}-R^{a}_{~bmn}e^{m}\h^{n}e^{b}
-F\h^{a}-F_{mn}e^{m}\h^{n}e^{a} \ , \\
\label{BRST-TWOFORM-R}
sR^{a}_{~b}\=(dR^{a}_{~bmn})e^{m}\h^{n}-R^{a}_{~bmn}e^{m}(d\h^{n})
+\th^{a}_{~c}R^{c}_{~b}-\th^{c}_{~b}R^{a}_{~c} \non
\-R^{a}_{~bmn}\o^{m}_{~k}e^{k}\h^{n}+\o^{a}_{~ck}\h^{k}R^{c}_{~b}
-\o^{c}_{~bk}\h^{k}R^{a}_{~c}+\o^{a}_{~c}R^{c}_{~bmn}e^{m}\h^{n} \non
\-\o^{c}_{~b}R^{a}_{~cmn}e^{m}\h^{n}+R^{a}_{~bmn}T^{m}\h^{n}
-R^{a}_{~bmn}Ae^{m}\h^{n} \ , \\
\label{BRST-TWOFORM-F}
sF\=(dF_{mn})e^{m}\h^{n}-F_{mn}e^{m}(d\h^{n})
+F_{mn}T^{m}\h^{n} \non
\-F_{mn}\o^{m}_{~k}e^{k}\h^{n}
-F_{mn}Ae^{m}\h^{n} \ .
\eea
To calculate out the corresponding zero forms
$(T^{a}_{mn}, R^{a}_{~bmn}, F_{mn})$ one needs the partial derivative
in the tangent space of these fields. From the Bianchi identities
\equ{BI} one gets the complete antisymmetrized relations
\bea
\label{PARTDERIV-T}
\6_{k}T^{a}_{mn}+\6_{m}T^{a}_{nk}+\6_{n}T^{a}_{km}\=
R^{a}_{~kmn}+R^{a}_{~mnk}+R^{a}_{~nkm} \non
&+&\d^{a}_{k}F_{mn}+\d^{a}_{m}F_{nk}+\d^{a}_{n}F_{km} \non
&-&\o^{a}_{~bk}T^{b}_{mn}-\o^{a}_{~bm}T^{b}_{nk}
-\o^{a}_{~bn}T^{b}_{km} \non
&+&A_{k}T^{a}_{mn}+A_{m}T^{a}_{nk}+A_{n}T^{a}_{km} \non
&-&T^{a}_{lk}T^{l}_{mn}-T^{a}_{lm}T^{l}_{nk}
-T^{a}_{ln}T^{l}_{km} \non
&+&T^{a}_{lk}\o^{l}_{~nm}+T^{a}_{ln}\o^{l}_{~mk}
+T^{a}_{lm}\o^{l}_{~kn} \non
&-&T^{a}_{lk}\o^{l}_{~mn}-T^{a}_{lm}\o^{l}_{~nk}
-T^{a}_{ln}\o^{l}_{~km} \ , \\
\label{PARTDERIV-R}
\6_{k}R^{a}_{~bmn}+\6_{m}R^{a}_{~bnk}+\6_{n}R^{a}_{~bkm}\=
-\o^{a}_{~ck}R^{c}_{~bmn}-\o^{a}_{~cm}R^{c}_{~bnk}
-\o^{a}_{~cn}R^{c}_{~bkm} \non
&+&\o^{c}_{~bk}R^{a}_{~cmn}+\o^{c}_{~bm}R^{a}_{~cnk}
+\o^{c}_{~bn}R^{a}_{~ckm} \non
&-&R^{a}_{~blk}T^{l}_{mn}-R^{a}_{~blm}T^{l}_{nk}
-R^{a}_{~bln}T^{l}_{km} \non
&+&R^{a}_{~blk}\o^{l}_{~nm}+R^{a}_{~bln}\o^{l}_{~mk}
+R^{a}_{~blm}\o^{l}_{~kn} \non
&-&R^{a}_{~blk}\o^{l}_{~mn}-R^{a}_{~blm}\o^{l}_{~nk}
-R^{a}_{~bln}\o^{l}_{~km} \non
&+&2A_{k}R^{a}_{~bmn}+2A_{m}R^{a}_{~bnk}+2A_{n}R^{a}_{~bkm} \ , \\
\label{PARTDERIV-F}
\6_{k}F_{mn}+\6_{m}F_{nk}+\6_{n}F_{km}\=
-F_{lk}T^{l}_{mn}-F_{lm}T^{l}_{nk}-F_{ln}T^{l}_{km} \non
&+&F_{lk}\o^{l}_{~nm}+F_{ln}\o^{l}_{~mk}+F_{lm}\o^{l}_{~kn} \non
&-&F_{lk}\o^{l}_{~mn}-F_{lm}\o^{l}_{~nk}-F_{ln}\o^{l}_{~km} \non
&+&2A_{k}F_{mn}+2A_{m}F_{nk}+2A_{n}F_{km} \ .
\eea
Inserting eqs.\equ{PARTDERIV-T}-\equ{PARTDERIV-F} into
eqs.\equ{BRST-TWOFORM-T}-\equ{BRST-TWOFORM-F} leads to the BRST
transformations of the zero form torsion field $T^{a}_{mn}$,
the zero form Riemann tensor $R^{a}_{~bmn}$, and the zero form
Weyl curvature $F_{mn}$
\bea
\label{BRST-ZERO-T}
sT^{a}_{mn}\=-\h^{k}\6_{k}T^{a}_{mn}+\th^{a}_{~b}T^{b}_{mn}
-\th^{k}_{~m}T^{a}_{kn}
-\th^{k}_{~n}T^{a}_{mk}-\s T^{a}_{mn} \ , \\
\label{BRST-ZERO-R}
sR^{a}_{~bmn}\=-\h^{k}\6_{k}R^{a}_{~bmn}+\th^{a}_{~c}R^{c}_{~bmn}
-\th^{c}_{~b}R^{a}_{~cmn}-\th^{k}_{~m}R^{a}_{~bkn} \non
\-\th^{k}_{~n}R^{a}_{~bmk}-2\s R^{a}_{~bmn} \ , \\
\label{BRST-ZERO-F}
sF_{mn}\=-\h^{k}\6_{k}F_{mn}-\th^{k}_{~m}F_{kn}
-\th^{k}_{~n}F_{mk}-2\s F_{mn} \ .
\eea
These equations, when rewritten in terms of the variable $\x^{\mu}$,
take the form (see eqs.\equ{BRST-TRANS})
\bea
sT^{a}_{\mu\nu}\=
-\x^{\l}\6_{\l}T^{a}_{\mu\nu}-(\6_{\mu}\x^{\l})T^{a}_{\l\nu}
-(\6_{\nu}\x^{\l})T^{a}_{\mu\l}
+\th^{a}_{~b}T^{b}_{\mu\nu}+\s T^{a}_{\mu\nu} \ , \non
sR^{a}_{~b\mu\nu}\=
-\x^{\l}\6_{\l}R^{a}_{~b\mu\nu}
-(\6_{\mu}\x^{\l})R^{a}_{~b\l\nu}
-(\6_{\nu}\x^{\l})R^{a}_{~b\mu\l}
+\th^{a}_{~c}R^{c}_{~b\mu\nu}
-\th^{c}_{~b}R^{a}_{~c\mu\nu} \ , \non
sF_{\mu\nu}\=-\x^{\l}\6_{\l}F_{\mu\nu}
-(\6_{\mu}\x^{\l})F_{\l\nu}-(\6_{\nu}\x^{\l})F_{\mu\l} \ .
\eea

Notice that, contrary to the case of the usual
space-time derivative $\6_{\mu}$, the operator $\6_{a}$ does not
commute with the BRST operator $s$ or with the exterior
derivative $d$ due to the explicit presence of the
vielbein $e^{a}$ (see Appendix A for the detailed calculations).
One has:
\bea
[s,\6_{m}]\=(\6_{m}\h^{k}-\th^{k}_{~m}-T^{k}_{mn}\h^{n}
-\o^{k}_{~mn}\h^{n}+\o^{k}_{~nm}\h^{n} \non
\+A_{m}\h^{k}-A_{n}\h^{n}\d^{k}_{m}-\s\d^{k}_{m})\6_{k} \ ,
\eea
and
\bea
[d,\6_{m}]\=(T^{k}_{mn}e^{n}+\o^{k}_{~mn}e^{n}
-\o^{k}_{~nm}e^{n}-A_{m}e^{k} \non
\+A_{n}e^{n}\d^{k}_{m}-(\6_{m}e^{k}))\6_{k} \ .
\eea
Also the commutator of two tangent space derivatives does not
vanish
\be
[\6_{m},\6_{n}]=-(T^{k}_{mn}+\o^{k}_{~mn}-\o^{k}_{~nm}
-A_{m}\d^{k}_{n}+A_{n}\d^{k}_{m})\6_{k} \ .
\ee
Nevertheless, taking into account the vielbein transformation
\equ{BRSTE}, one consistently verifies that
\be
\{s,d\}=0~~~,~~~d^{2}=0~~~,~~~s^{2}=0 \ .
\ee


\subsection{BRST transformations and Bianchi identities}

The last section is devoted to collect, as a summary for the reader,
the whole set of BRST transformations and the Bianchi identities which
emerge from the Maurer-Cartan horizontality conditions
\equ{MCG1}-\equ{MCG5} and from eqs.\equ{DEFTWO}, \equ{GBI} for
each form sector and ghost number.

\begin{itemize}

\item {\bf Form sector two, ghost number zero $(T^{a}, R^{a}_{~b}, F)$}

\bea
\label{FORMTWO}
sT^{a}\=(dT^{a}_{mn})e^{m}\h^{n}-T^{a}_{mn}e^{m}d\h^{n}
+\th^{a}_{~b}T^{b}+\s T^{a}+T^{a}_{mn}T^{m}\h^{n} \non
\-T^{a}_{mn}\o^{m}_{~k}e^{k}\h^{n}+\o^{a}_{~bk}\h^{k}T^{b}
+\o^{a}_{~b}T^{b}_{mn}e^{m}\h^{n}+A_{k}\h^{k}T^{a} \non
\-R^{a}_{~b}\h^{b}-R^{a}_{~bmn}e^{m}\h^{n}e^{b}
-F\h^{a}-F_{mn}e^{m}\h^{n}e^{a} \ , \non
sR^{a}_{~b}\=(dR^{a}_{~bmn})e^{m}\h^{n}-R^{a}_{~bmn}e^{m}d\h^{n}
+\th^{a}_{~c}R^{c}_{~b}-\th^{c}_{~b}R^{a}_{~c} \non
\-R^{a}_{~bmn}\o^{m}_{~k}e^{k}\h^{n}+\o^{a}_{~ck}\h^{k}R^{c}_{~b}
-\o^{c}_{~bk}\h^{k}R^{a}_{~c}+\o^{a}_{~c}R^{c}_{~bmn}e^{m}\h^{n} \non
\-\o^{c}_{~b}R^{a}_{~cmn}e^{m}\h^{n}+R^{a}_{~bmn}T^{m}\h^{n}
-R^{a}_{~bmn}Ae^{m}\h^{n} \ , \non
sF\=(dF_{mn})e^{m}\h^{n}-F_{mn}e^{m}d\h^{n}
+F_{mn}T^{m}\h^{n} \non
\-F_{mn}\o^{m}_{~k}e^{k}\h^{n}
-F_{mn}Ae^{m}\h^{n} \ .
\eea
For the Bianchi identities one has
\bea
\label{BIG}
&&DT^{a}=dT^{a}+\o^{a}_{~b}T^{b}+AT^{a}
=R^{a}_{~b}e^{b}+Fe^{a} \ , \non
&&DR^{a}_{~b}=dR^{a}_{~b}+\o^{a}_{~c}R^{c}_{~b}
-\o^{c}_{~b}R^{a}_{~c}=0 \ , \non
&&DF=dF=0 \ .
\eea

\item {\bf Form sector one, ghost number zero $(e^{a}, \o^{a}_{~b}, A)$}

\bea
\label{FORM1}
se^{a}\=d\h^{a}+\o^{a}_{~b}\h^{b}+\th^{a}_{~b}e^{b}
+\o^{a}_{~bm}\h^{m}e^{b}+A_{m}\h^{m}e^{a}+\s e^{a} \non
\+A\h^{a}-T^{a}_{mn}e^{m}\h^{n} \ , \non
s\o^{a}_{~b}\=(d\o^{a}_{~bm})\h^{m}+\o^{a}_{~bm}d\h^{m}
+d\th^{a}_{~b}+\o^{a}_{~cm}\h^{m}\o^{c}_{~b}
+\th^{a}_{~c}\o^{c}_{~b} \non
\-\o^{c}_{~bm}\h^{m}\o^{a}_{~c}
-\th^{c}_{~b}\o^{a}_{~c}-R^{a}_{~bmn}e^{m}\h^{n} \ , \non
sA\=(dA_{m})\h^{m}+A_{m}d\h^{m}+d\s-F_{mn}e^{m}\h^{n} \ .
\eea
The exterior derivatives of these fields are given by the
definitions of the two form field strengths \equ{TWO-FORMS}
\bea
&&d\o^{a}_{~b}=R^{a}_{~b}-\o^{a}_{~c}\o^{c}_{~b} \ , \non
&&de^{a}=T^{a}-\o^{a}_{~b}e^{b}-Ae^{a} \ , \non
&&dA=F \ .
\eea

\item {\bf Form sector zero, ghost number zero
$(\ph, \o^{a}_{~bm}, A_{m}, T^{a}_{mn}, R^{a}_{~bmn}, F_{mn})$ }

\bea
s\ph\=-\h^{m}\6_{m}\ph-\frac{N-2}{2}\s\ph \ , \non
s\o^{a}_{~bm}\=-\h^{k}\6_{k}\o^{a}_{~bm}-\6_{m}\th^{a}_{~b}
+\th^{a}_{~c}\o^{c}_{~bm}-\th^{c}_{~b}\o^{a}_{~cm}
-\th^{k}_{~m}\o^{a}_{~bk}-\s\o^{a}_{~bm} \ , \non
sA_{m}\=-\h^{k}\6_{k}A_{m}-\th^{k}_{~m}A_{k}-\6_{m}\s-A_{m}\s \ , \non
sT^{a}_{mn}\=-\h^{k}\6_{k}T^{a}_{mn}+\th^{a}_{~b}T^{b}_{mn}
-\th^{k}_{~m}T^{a}_{kn}
-\th^{k}_{~n}T^{a}_{mk}-\s T^{a}_{mn} \ , \non
sR^{a}_{~bmn}\=-\h^{k}\6_{k}R^{a}_{~bmn}+\th^{a}_{~c}R^{c}_{~bmn}
-\th^{c}_{~b}R^{a}_{~cmn}-\th^{k}_{~m}R^{a}_{~bkn} \non
\-\th^{k}_{~n}R^{a}_{~bmk}-2\s R^{a}_{~bmn} \ , \non
sF_{mn}\=-\h^{k}\6_{k}F_{mn}-\th^{k}_{~m}F_{kn}
-\th^{k}_{~n}F_{mk}-2\s F_{mn} \ .
\eea
The Bianchi identities \equ{BIG} are projected on the zero form
torsion $T^{a}_{mn}$, the zero form Riemann tensor $R^{a}_{~bmn}$,
and on the zero form Weyl curvature $F_{mn}$ to give
\bea
\label{PRO-BIANCHI}
dT^{a}_{mn}\=(\6_{k}T^{a}_{mn})e^{k} \non
\=(R^{a}_{~kmn}+R^{a}_{~mnk}+R^{a}_{~nkm} \non
\+\d^{a}_{k}F_{mn}+\d^{a}_{m}F_{nk}+\d^{a}_{n}F_{km} \non
\-\o^{a}_{~bk}T^{b}_{mn}-\o^{a}_{~bm}T^{b}_{nk}
-\o^{a}_{~bn}T^{b}_{km} \non
\+A_{k}T^{a}_{mn}+A_{m}T^{a}_{nk}+A_{n}T^{a}_{km} \non
\-T^{a}_{lk}T^{l}_{mn}-T^{a}_{lm}T^{l}_{nk}
-T^{a}_{ln}T^{l}_{km} \non
\+T^{a}_{lk}\o^{l}_{~nm}+T^{a}_{ln}\o^{l}_{~mk}
+T^{a}_{lm}\o^{l}_{~kn} \non
\-T^{a}_{lk}\o^{l}_{~mn}-T^{a}_{lm}\o^{l}_{~nk}
-T^{a}_{ln}\o^{l}_{~km} \non
\-\6_{m}T^{a}_{nk}-\6_{n}T^{a}_{km})e^{k} \ , \non
dR^{a}_{~bmn}\=(\6_{k}R^{a}_{~bmn})e^{k} \non
\=(-\o^{a}_{~ck}R^{c}_{~bmn}-\o^{a}_{~cm}R^{c}_{~bnk}
-\o^{a}_{~cn}R^{c}_{~bkm} \non
\+\o^{c}_{~bk}R^{a}_{~cmn}+\o^{c}_{~bm}R^{a}_{~cnk}
+\o^{c}_{~bn}R^{a}_{~ckm} \non
\-R^{a}_{~blk}T^{l}_{mn}-R^{a}_{~blm}T^{l}_{nk}
-R^{a}_{~bln}T^{l}_{km} \non
\+R^{a}_{~blk}\o^{l}_{~nm}+R^{a}_{~bln}\o^{l}_{~mk}
+R^{a}_{~blm}\o^{l}_{~kn} \non
\-R^{a}_{~blk}\o^{l}_{~mn}-R^{a}_{~blm}\o^{l}_{~nk}
-R^{a}_{~bln}\o^{l}_{~km} \non
\+2A_{k}R^{a}_{~bmn}+2A_{m}R^{a}_{~bnk}+2A_{n}R^{a}_{~bkm} \non
\-\6_{m}R^{a}_{~bnk}-\6_{n}R^{a}_{~bkm})e^{k} \ , \non
dF_{mn}\=(\6_{k}F_{mn})e^{k} \non
\=(-F_{lk}T^{l}_{mn}-F_{lm}T^{l}_{nk}-F_{ln}T^{l}_{km} \non
\+F_{lk}\o^{l}_{~nm}+F_{ln}\o^{l}_{~mk}+F_{lm}\o^{l}_{~kn} \non
\-F_{lk}\o^{l}_{~mn}-F_{lm}\o^{l}_{~nk}-F_{ln}\o^{l}_{~km} \non
\+2A_{k}F_{mn}+2A_{m}F_{nk}+2A_{n}F_{km} \non
\-\6_{m}F_{nk}-\6_{n}F_{km})e^{k} \ .
\eea
One has also the equations
\bea
d\o^{a}_{~bm}\=(\6_{n}\o^{a}_{~bm})e^{n} \non
\=-(R^{a}_{~bmn}
-\o^{a}_{~cm}\o^{c}_{~bn}+\o^{a}_{~cn}\o^{c}_{~bm} \non
\-\o^{a}_{~bk}T^{k}_{mn}+\o^{a}_{~bk}\o^{k}_{~nm}
-\o^{a}_{~bk}\o^{k}_{~mn} \non
\+\o^{a}_{~bn}A_{m}-\o^{a}_{~bm}A_{n}
-\6_{m}\o^{a}_{~bn})e^{n} \ , \non
dA_{m}\=(\6_{n}A_{m})e^{n} \non
\=-(F_{mn}-A_{k}T^{k}_{mn}-A_{k}\o^{k}_{~mn} \non
\+A_{k}\o^{k}_{~nm}-\6_{m}A_{n})e^{n} \ .
\eea

\item {\bf Form sector zero, ghost number one $(\h^{a}, \th^{a}_{~b}, \s)$ }

The BRST transformations for the ghost fields are given by
\bea
\label{FORMZERO}
s\h^{a}\=\th^{a}_{~b}\h^{b}+\o^{a}_{~bm}\h^{m}\h^{b}
+A_{m}\h^{m}\h^{a}+\s\h^{a}
-\frac{1}{2}T^{a}_{mn}\h^{m}\h^{n} \ , \non
s\th^{a}_{~b}\=\th^{a}_{~c}\th^{c}_{~b}
-\h^{k}\6_{k}\th^{a}_{~b} \ , \non
s\s\=-\h^{k}\6_{k}\s \ .
\eea

\item {\bf Commutator relations for the tangent space
derivative $\6_{m}$ }

The following commutator relations are valid:
\bea
\label{COMMUT-1}
&&[s,\6_{m}]=(\6_{m}\h^{k}-\th^{k}_{~m}-T^{k}_{mn}\h^{n}
-\o^{k}_{~mn}\h^{n}+\o^{k}_{~nm}\h^{n} \non
&&~~~~~~~~~+~A_{m}\h^{k}-A_{n}\h^{n}\d^{k}_{m}
-\s\d^{k}_{m})\6_{k} \ , \non
&&[d,\6_{m}]=(T^{k}_{mn}e^{n}+\o^{k}_{~mn}e^{n}
-\o^{k}_{~nm}e^{n} \non
&&~~~~~~~~~-~A_{m}e^{k}+A_{n}e^{n}\d^{k}_{m}
-(\6_{m}e^{k}))\6_{k} \ ,
\eea
and
\be
\label{COMMUT-2}
[\6_{m},\6_{n}]=-(T^{k}_{mn}+\o^{k}_{~mn}-\o^{k}_{~nm}
-A_{m}\d^{k}_{n}+A_{n}\d^{k}_{m})\6_{k} \ .
\ee
\makebox[1cm]{}

\item {\bf Algebra between $s$ and $d$ }

{}From the above transformations it follows:
\be
s^{2}=0 \ ,~~~d^{2}=0 \ ,
\ee
and
\be
\{s,d\}=0 \ .
\ee

\end{itemize}

Let us conclude this section by making two remarks. The first one
concerns the role played by the torsion $T$ in the
BRST transformations. We emphasize that, as one
can see from eqs.\equ{FORMTWO}-\equ{FORMZERO}, a fully tangent space
formulation of the gravitational algebra can be obtained only
when the torsion is explicitly present.

The second remark is
related to the use of the variable $\h^a$. Observe
that, when expressed in terms of $\h^a$, the BRST transformation
of the vielbein $e^a$ in \equ{FORM1} starts with a term linear
in the fields (i.e. the term $d\h^a$). This feature makes
the analogy between gravitational and gauge theories
more transparent.
Moreover, it suggests that one
may compute the local cohomology of the gravitational
BRST operator $s$~\cite{next}
without expanding the vielbein $e^a$ around a flat
background, as shown in~\cite{dragon}.


\section{Descent equations and decomposition}

The discussion of invariant Lagrangians and anomalies implies
to find non-trivial solutions of the
so-called Wess-Zumino consistency condition~\cite{wess}
formulated in terms of the BRST transformations
\be
\label{WESS-ZUMINO}
sa=0 \ ,
\ee
where $s$ is the nilpotent BRST operator and $a$ is an integrated
local field polynomial in the space of differential forms~\cite{dviolette}.
The use of the space of polynomials of forms is not a restriction
on the generality of the solutions of the consistency equation,
as recently proven by M. Dubois-Violette et al.~\cite{dviolette}.
Non-trivial solutions of \equ{WESS-ZUMINO}
are given by the descent-equations
technique~\cite{dragon,baulieu,dviolette,stora,brandt,zumino2}.
Setting
\be
a=\int\AA \ ,
\ee
condition \equ{WESS-ZUMINO} translates into the local
equation
\be
\label{LOCAL-EQ}
s\AA+d\hat\AA=0 \ ,
\ee
where $\AA$ are some local polynomials and $d=dx^{\mu}\6_{\mu}$
denotes the exterior space-time derivative.
$\AA$ is said non-trivial if
\be
\AA \not= s\BB+d\hat\BB \ ,
\ee
with $\BB$ and $\hat\BB$ local polynomials.
The volume form ($N$-form) $\AA$ with a given ghost number $G$
is denoted by $\AA^{G}_{N}$. The local equation \equ{LOCAL-EQ} reads then
\be
\label{LOCAL-EQUATION}
s\AA^{G}_{N}+d\AA^{G+1}_{N-1}=0 \ .
\ee
In this case the integral of $\AA$ on space-time, $\int\AA$,
identifies a cohomology class of the BRST operator $s$ and, according
to its ghost number $G$, it corresponds to an invariant Lagrangian
($G=0$) or to an anomaly ($G=1$).

By applying the BRST operator $s$ on the local
equation \equ{LOCAL-EQUATION}, due to the relations
\be
s^{2}=d^{2}=sd+ds=0
\ee
and to the algebraic
Poincar\'e Lemma~\cite{cotta,dragon}
\be
\label{POINCARE-LEMMA}
d\O=0~~ \Longleftrightarrow ~~\O=d\widehat{\O}+d^{N}\!x~\LL+const \ ,
\ee
it follows for the next cocycle
\be
\label{COC-1}
s\AA^{G+1}_{N-1}+d\AA^{G+2}_{N-2}=0 \ ,
\ee
because $s\AA^{G+1}_{N-1}$ is not a volume form nor a constant form.
It is easily seen that repeated applications of the operator $s$
generate a tower of descent equations
\bea
\label{LADDER-2}
&&s\AA^{G}_{N}+d\AA^{G+1}_{N-1}=0 \ ,\non
&&s\AA^{G+1}_{N-1}+d\AA^{G+2}_{N-2}=0 \ ,\non
&&~~~~~.....\non
&&~~~~~.....\non
&&s\AA^{G+N-1}_{1}+d\AA^{G+N}_{0}=0 \ ,\non
&&s\AA^{G+N}_{0}=0 \ ,
\eea
which ends with a zero form cocycle $\AA^{G+N}_{0}$.

The main idea to solve the tower of descent equations \equ{LADDER-2}
is based on the fact that the exterior derivative $d$ can be written
as a BRST commutator in the following sense
\be
\label{BRST-COMM}
d=-[s,\d] \ ,
\ee
where $\d$ is an operator which will be specified later. In general
this operator decrease the ghost number $g$ by one and increase the form
degree $p$ by one
\be
\d\AA^{g}_{p}=\AA^{g-1}_{p+1} \ .
\ee


\subsection{Pure Yang-Mills}

In order to elaborate the general idea of solving the tower
\equ{LADDER-2} one starts again with the discussion of the pure
Yang-Mills theory, as it was done in a recent paper
of S.P. Sorella~\cite{silvio}.
For this one introduces the operators $\d$ and $\GG$ defined by
\be
\label{DEC-YM}
\d=-A^{a}\frac{\6}{\6 c^{a}}+(F^{a}+\frac{1}{2}f^{abc}A^{b}A^{c})
\frac{\6}{\6 dc^{a}} \ ,
\ee
and
\be
\GG=-F^{a}\frac{\6}{\6 c^{a}}+f^{abc}F^{b}A^{c}\frac{\6}{\6 dc^{a}} \ .
\ee
It is easily verified that $\d$ and $\GG$ are respectively of degree
zero and one and that the following algebraic relations hold:
\be
\label{DECOMP-YM}
d=-[s,\d] \ ,
\ee
\be
\label{G-COMMUTATOR}
2\GG=[d,\d] \ ,
\ee
\be
\{s,\GG\}=0~~~~~,~~~~~\GG=0 \ ,
\ee
\be
\{d,\GG\}=0~~~~~,~~~~~[\GG,\d]=0 \ .
\ee
Notice that the closure of the algebra between $d$, $s$ and $\d$
requires the introduction of the operator $\GG$.
As an example, we discuss the three-dimensional Chern-Simons term,
which is also relevant for the two-dimensional gauge anomaly.
In this case the descent equations read:
\bea
\label{DESCENT-YM}
&&s\O^{0}_{3}+d\O^{1}_{2}=0 \ ,\non
&&s\O^{1}_{2}+d\O^{2}_{1}=0 \ ,\non
&&s\O^{2}_{1}+d\O^{3}_{0}=0 \ ,\non
&&s\O^{3}_{0}=0 \ ,
\eea
where $\O^{3}_{0}$ is the BRST invariant ghost monomial defined by
\be
\O^{3}_{0}=\frac{1}{3!}f^{abc}c^{a}c^{b}c^{c} \ .
\ee
Acting with the operator $\d$ of eq.\equ{DEC-YM} on the last equation
of the tower \equ{DESCENT-YM} one gets
\be
[\d,s]\O^{3}_{0}+s\d\O^{3}_{0}=0 \ ,
\ee
which, using the decomposition \equ{DECOMP-YM}, becomes
\be
\label{COCYCLE-1}
s\d\O^{3}_{0}+d\O^{3}_{0}=0 \ .
\ee
This equation shows that $\d\O^{3}_{0}$ gives a solution for the
cocycle $\O^{2}_{1}$ in the tower \equ{DESCENT-YM}.
Acting again with $\d$ on the eq.\equ{COCYCLE-1} and using the
algebraic relations \equ{DECOMP-YM},\equ{G-COMMUTATOR} one has
\be
\label{COCYC-2}
s\frac{\d\d}{2}\O^{3}_{0}-\GG\O^{3}_{0}+d\d\O^{3}_{0}=0 \ .
\ee
Moreover, with the relations
\bea
\GG\O^{3}_{0}\=s\widehat{\O}^{1}_{2} \ , \non
\widehat{\O}^{1}_{2}\=F^{a}c^{a} \ ,
\eea
eq.\equ{COCYC-2} can be rewritten as:
\be
\label{COCYCLE-2}
s(\frac{\d\d}{2}\O^{3}_{0}-\widehat{\O}^{1}_{2})+d\d\O^{3}_{0}=0 \ .
\ee
One sees that $(\frac{\d\d}{2}\O^{3}_{0}-\widehat{\O}^{1}_{2})$
gives a solution for $\O^{1}_{2}$ modulo trivial contributions.
To solve completely the tower
\equ{DESCENT-YM} one have to apply once more the operator $\d$
on the eq.\equ{COCYCLE-2}. After a little algebra one gets:
\be
\label{COCYCLE-3}
s(\frac{\d\d\d}{3!}\O^{3}_{0}-\d\widehat{\O}^{1}_{2})
+d(\frac{\d\d}{2}\O^{3}_{0}-\widehat{\O}^{1}_{2})=0 \ ,
\ee
which shows that the cocycle $\O^{0}_{3}$ can be identified with
$(\frac{\d\d\d}{3!}\O^{3}_{0}-\d\widehat{\O}^{1}_{2})$.
It is apparent then how repeated applications of the operator $\d$
on the zero form cocycle $\O^{3}_{0}$ and the use of the
operator $\GG$ in the tower give
in a simple way a solution of the descent equations.
Summarizing, the solution of the descent equations \equ{DESCENT-YM}
is given by
\bea
\O^{0}_{3}\=\frac{1}{3!}\d\d\d\O^{3}_{0}-\d\widehat{\O}^{1}_{2} \ , \\
\O^{1}_{2}\=\frac{1}{2}\d\d\O^{3}_{0}-\widehat{\O}^{1}_{2} \ , \\
\O^{2}_{1}\=\d\O^{3}_{0} \ ,
\eea
where
\bea
\widehat{\O}^{1}_{2}\=F^{a}c^{a} \ , \\
\O^{3}_{0}\=\frac{1}{3!}f^{abc}c^{a}c^{b}c^{c} \ , \\
\O^{2}_{1}\=-\frac{1}{2}f^{abc}A^{a}c^{b}c^{c}
=(dc^{a})c^{a}-s(A^{a}c^{a}) \ , \\
\label{ANOMALY-YM}
\O^{1}_{2}\=\frac{1}{2}f^{abc}A^{a}A^{b}c^{c}-F^{a}c^{a}
=-(dA^{a})c^{a} \ , \\
\label{CHERN-SIMONS-YM}
\O^{0}_{3}\=F^{a}A^{a}-\frac{1}{3!}f^{abc}A^{a}A^{b}A^{c} \ .
\eea
One sees then, that \equ{ANOMALY-YM} and \equ{CHERN-SIMONS-YM}
give respectively the two-dimensional gauge anomaly
(modulo a $d$-coboundary) and the three-dimensional
Chern-Simons term.


\subsection{Gravity with torsion}

Now one can apply the same technique to solve with an appropriate
decomposition \equ{DECOMP} the ladder for the gravitational case.
In a first step one neglects the Weyl symmetry to simplify the matter.
For this purpose one defines the operator $\d$ as
\be
\label{DECETA-0}
\d=-e^{a}\frac{\d}{\d\h^{a}} \ ,
\ee
or in terms of the basic fields
\bea
\label{DECETA}
\d\h^{a}\=-e^{a} \ ,\non
\d\Ph\=0~~~{\rm for}~~~\Ph=(\ph, \o, e, R, T, \th) \ .
\eea
It is easy to verify that $\d$ is of degree zero and that, together
with the BRST operator $s$, it obeys the following algebraic relations:
\be
\label{DEC}
[s,\d]=-d \ ,
\ee
and
\be
\label{EXCOM}
[d,\d]=0 \ .
\ee
One sees from eq.\equ{DEC} that the operator $\d$ allows to decompose
the exterior derivative $d$ as a BRST commutator.
This property, as already shown in~\cite{silvio}, gives an elegant and
simple procedure for solving the descent equations \equ{LADDER-2}.

The BRST transformations for gravity with torsion but without
Weyl transformations are
easily found by setting the corresponding
Weyl fields, namely the Weyl gauge field $A$, the Weyl ghost $\s$,
and the Weyl curvature $F$, to zero. Furthermore, for this case one has
also to modify the projected Bianchi identities \equ{PRO-BIANCHI}
and the commutator relations \equ{COMMUT-1}-\equ{COMMUT-2}.
This leads to the following summarized result:

\begin{itemize}

\item {\bf Form sector two, ghost number zero $(T^{a}, R^{a}_{~b})$}

\bea
\label{FORMTWO-GRAV}
sT^{a}\=(dT^{a}_{mn})e^{m}\h^{n}-T^{a}_{mn}e^{m}d\h^{n}
+\th^{a}_{~b}T^{b}+T^{a}_{mn}T^{m}\h^{n} \non
\-T^{a}_{mn}\o^{m}_{~k}e^{k}\h^{n}+\o^{a}_{~bk}\h^{k}T^{b}
+\o^{a}_{~b}T^{b}_{mn}e^{m}\h^{n} \non
\-R^{a}_{~b}\h^{b}-R^{a}_{~bmn}e^{m}\h^{n}e^{b} \ , \non
sR^{a}_{~b}\=(dR^{a}_{~bmn})e^{m}\h^{n}-R^{a}_{~bmn}e^{m}d\h^{n}
+\th^{a}_{~c}R^{c}_{~b}-\th^{c}_{~b}R^{a}_{~c} \non
\-R^{a}_{~bmn}\o^{m}_{~k}e^{k}\h^{n}+\o^{a}_{~ck}\h^{k}R^{c}_{~b}
-\o^{c}_{~bk}\h^{k}R^{a}_{~c}+\o^{a}_{~c}R^{c}_{~bmn}e^{m}\h^{n} \non
\-\o^{c}_{~b}R^{a}_{~cmn}e^{m}\h^{n}+R^{a}_{~bmn}T^{m}\h^{n} \ .
\eea
For the Bianchi identities one has
\bea
\label{BIG-GRAV}
&&DT^{a}=dT^{a}+\o^{a}_{~b}T^{b}=R^{a}_{~b}e^{b} \ , \non
&&DR^{a}_{~b}=dR^{a}_{~b}+\o^{a}_{~c}R^{c}_{~b}-\o^{c}_{~b}R^{a}_{~c}=0 \ .
\eea

\item {\bf Form sector one, ghost number zero $(e^{a}, \o^{a}_{~b})$}

\bea
se^{a}\=d\h^{a}+\o^{a}_{~b}\h^{b}+\th^{a}_{~b}e^{b}
+\o^{a}_{~bm}\h^{m}e^{b}
-T^{a}_{mn}e^{m}\h^{n} \ , \non
s\o^{a}_{~b}\=(d\o^{a}_{~bm})\h^{m}+\o^{a}_{~bm}d\h^{m}
+d\th^{a}_{~b}+\o^{a}_{~cm}\h^{m}\o^{c}_{~b}
+\th^{a}_{~c}\o^{c}_{~b} \non
\-\o^{c}_{~bm}\h^{m}\o^{a}_{~c}
-\th^{c}_{~b}\o^{a}_{~c}-R^{a}_{~bmn}e^{m}\h^{n} \ .
\eea

The exterior derivatives of these fields are given by the defintions
of the two form field strengths
\bea
&&de^{a}=T^{a}-\o^{a}_{~b}e^{b} \ , \non
&&d\o^{a}_{~b}=R^{a}_{~b}-\o^{a}_{~c}\o^{c}_{~b} \ .
\eea

\item {\bf Form sector zero, ghost number zero
$(\ph, \o^{a}_{~bm}, R^{a}_{~bmn}, T^{a}_{mn})$ }

\bea
s\ph\=-\h^{m}\6_{m}\ph \ , \non
s\o^{a}_{~bm}\=-\h^{k}\6_{k}\o^{a}_{~bm}-\6_{m}\th^{a}_{~b}
+\th^{a}_{~c}\o^{c}_{~bm}-\th^{c}_{~b}\o^{a}_{~cm}
-\th^{k}_{~m}\o^{a}_{~bk} \ , \non
sT^{a}_{mn}\=-\h^{k}\6_{k}T^{a}_{mn}+\th^{a}_{~k}T^{k}_{mn}
-\th^{k}_{~m}T^{a}_{kn}-\th^{k}_{~n}T^{a}_{mk} \ , \non
sR^{a}_{~bmn}\=-\h^{k}\6_{k}R^{a}_{~bmn}+\th^{a}_{~c}R^{c}_{~bmn}
-\th^{c}_{~b}R^{a}_{~cmn}-\th^{k}_{~m}R^{a}_{~bkn}
-\th^{k}_{~n}R^{a}_{~bmk} \ .
\eea
{}From the equations \equ{PRO-BIANCHI} one gets
\bea
dT^{a}_{mn}\=(\6_{k}T^{a}_{mn})e^{k} \non
\=(R^{a}_{~kmn}+R^{a}_{~mnk}+R^{a}_{~nkm} \non
\-\o^{a}_{~bk}T^{b}_{mn}-\o^{a}_{~bm}T^{b}_{nk}
-\o^{a}_{~bn}T^{b}_{km} \non
\-T^{a}_{lk}T^{l}_{mn}-T^{a}_{lm}T^{l}_{nk}
-T^{a}_{ln}T^{l}_{km} \non
\+T^{a}_{lk}\o^{l}_{~nm}+T^{a}_{ln}\o^{l}_{~mk}
+T^{a}_{lm}\o^{l}_{~kn} \non
\-T^{a}_{lk}\o^{l}_{~mn}-T^{a}_{lm}\o^{l}_{~nk}
-T^{a}_{ln}\o^{l}_{~km} \non
\-\6_{m}T^{a}_{nk}-\6_{n}T^{a}_{km})e^{k} \ , \non
dR^{a}_{~bmn}\=(\6_{k}R^{a}_{~bmn})e^{k} \non
\=(-\o^{a}_{~ck}R^{c}_{~bmn}-\o^{a}_{~cm}R^{c}_{~bnk}
-\o^{a}_{~cn}R^{c}_{~bkm} \non
\+\o^{c}_{~bk}R^{a}_{~cmn}+\o^{c}_{~bm}R^{a}_{~cnk}
+\o^{c}_{~bn}R^{a}_{~ckm} \non
\-R^{a}_{~blk}T^{l}_{mn}-R^{a}_{~blm}T^{l}_{nk}
-R^{a}_{~bln}T^{l}_{km} \non
\+R^{a}_{~blk}\o^{l}_{~nm}+R^{a}_{~bln}\o^{l}_{~mk}
+R^{a}_{~blm}\o^{l}_{~kn} \non
\-R^{a}_{~blk}\o^{l}_{~mn}-R^{a}_{~blm}\o^{l}_{~nk}
-R^{a}_{~bln}\o^{l}_{~km} \non
\-\6_{m}R^{a}_{~bnk}-\6_{n}R^{a}_{~bkm})e^{k} \ .
\eea
In addition, one has also the equation
\bea
d\o^{a}_{~bm}\=(\6_{n}\o^{a}_{~bm})e^{n}\non
\=(-R^{a}_{~bmn}+\o^{a}_{~cm}\o^{c}_{~bn}-\o^{a}_{~cn}\o^{c}_{~bm}\non
\+\o^{a}_{~bk}T^{k}_{mn}-\o^{a}_{~bk}\o^{k}_{~nm}
+\o^{a}_{~bk}\o^{k}_{~mn}+\6_{m}\o^{a}_{~bn})e^{n} \ .
\eea

\item {\bf Form sector zero, ghost number one $(\th^{a}_{~b}, \h^{a})$ }

\bea
\label{FORMZERO-GRAV}
s\h^{a}\=\th^{a}_{~b}\h^{b}+\o^{a}_{~bm}\h^{m}\h^{b}
-\frac{1}{2}T^{a}_{mn}\h^{m}\h^{n} \ , \non
s\th^{a}_{~b}\=\th^{a}_{~c}\th^{c}_{~b}
-\h^{k}\6_{k}\th^{a}_{~b} \ .
\eea

\item {\bf Commutator relations for the tangent space
derivative $\6_{m}$ }

The following commutator relations are valid:
\bea
\label{COMMUT-1-GRAV}
&&[s,\6_{m}]=(\6_{m}\h^{k}-\th^{k}_{~m}-T^{k}_{mn}\h^{n}
-\o^{k}_{~mn}\h^{n}+\o^{k}_{~nm}\h^{n})\6_{k} \ , \non
&&[d,\6_{m}]=(T^{k}_{mn}e^{n}+\o^{k}_{~mn}e^{n}
-\o^{k}_{~nm}e^{n}
-(\6_{m}e^{k}))\6_{k} \ ,
\eea
and
\be
\label{COMMUT-2-GRAV}
[\6_{m},\6_{n}]=-(T^{k}_{mn}+\o^{k}_{~mn}-\o^{k}_{~nm})\6_{k} \ .
\ee
\makebox[1cm]{}

\item {\bf Algebra between $s$ and $d$ }

{}From the above transformations it follows:
\be
s^{2}=0 \ ,~~~d^{2}=0 \ ,
\ee
and
\be
\{s,d\}=0 \ .
\ee

\end{itemize}


\subsection{Decomposition in the presence of Weyl symmetry}

This section is dedicated to the description of the generalization
including also the Weyl symmetry.
Analogous to eq.\equ{DECETA} one introduces the operator $\d$ defined as
\be
\label{WEYL-DECETA-0}
\d=-e^{a}\frac{\d}{\d\h^{a}} \ ,
\ee
or in terms of the basic fields
\bea
\label{WEYL-DECETA}
\d\h^{a}\=-e^{a} \ ,\non
\d\Ph\=0~~~{\rm for}~~~\Ph=(\ph, \o, e, A, R, T, F, \th, \s) \ .
\eea
It is easy to verify that $\d$ is of degree zero and that, together
with the BRST operator $s$, it obeys the same algebraic relations
as in the case without Weyl transformations\footnote{Remark that contrary
to the previous section now $s$ denotes the full BRST operator
defined in Section 3.3.}:
\be
\label{WEYL-DEC}
[s,\d]=-d \ ,
\ee
and
\be
\label{WEYL-EXCOM}
[d,\d]=0 \ .
\ee
As in the case for gravity with torsion and without
Weyl symmetry the algebra is unchanged and
the operator $\d$ allows again the decomposition of
the exterior derivative $d$ as a BRST commutator.

In order to summarize the meaning of the $\d$-operator
one has to take into consideration the following points:

\begin{itemize}

\item {}

In the discussion of non-abelian gauge anomalies without gravity
one is enforced to introduce also an additional operator $\GG$ in order
to have a closed algebra between the operators $d$, $s$, $\d$, and $\GG$.
This operator, already present in the work of Brandt et al.~\cite{brandt},
generates together with the BRST operator $s$ a new tower of descent
equations which are easily disentangled by using the general results of
the cohomology of $s$~\cite{silvio}.

\item {}

The decomposition without the operator $\GG$ is also present in topological
field theories. In the Chern-Simons theory in three space-time
dimensions (quantized in the Landau gauge or in the axial gauge) the operator
$\d$ corresponds to a {\it new linear vector-supersymmetry}, which allows
to prove the finiteness of the underlying model in a very elegant
manner~\cite{gieres,birmingham2,delduc,lucchesi,brandhuber}.
This linear supersymmetry is also present in string and superstring theory
as it was shown recently~\cite{mauricio,boresch}.

\item {}

In our present considerations we have now shown that in gravity with torsion
(with or without Weyl symmetry) the $\GG$-operator is again absent.
The meaning of the operator $\GG$ is, up to our knowledge,
not yet fully understood.
More recently, we have some hints that also in topological Yang-Mills theory
in four space-time dimensions the operator $\d$ corresponds again to a
linear vector-supersymmetry, with a vanishing $\GG$-operator~\cite{next}.

\end{itemize}


\subsection{General solution of the tower}

Now one solves explicitely the tower of descent equations
for the gravitational case with and without Weyl symmetry. Therefore, without
Weyl symmetry, we expect that for ghost number $G$ and form degree $N$,
$N$ being the dimension of the space-time,
the $\O^{G}_{N}$ are local polynomials in the
fields $(\ph, e^{a}, \o^{a}_{~bm}, T^{a}_{mn}, R^{a}_{~bmn}
\h^{a}, \th^{a}_{~b})$ and their derivatives, whereas for the case with
Weyl symmetry one has to add the Weyl gauge field $A_{m}$, the corresponding
Weyl ghost $\s$, the Weyl curvature $F_{mn}$, and derivatives of them.
The tower of descent equations is given by
\bea
\label{TOWER}
&&s\O^{G}_{N}+d\O^{G+1}_{N-1}=0 \ ,\non
&&s\O^{G+1}_{N-1}+d\O^{G+2}_{N-2}=0 \ ,\non
&&~~~~~.....\non
&&~~~~~.....\non
&&s\O^{G+N-1}_{1}+d\O^{G+N}_{0}=0 \ ,\non
&&s\O^{G+N}_{0}=0 \ ,
\eea
with $(\O^{G+1}_{N-1},...., \O^{G+N-1}_{1}, \O^{G+N}_{0})$ local
polynomials which, without loss of generality, will be always
considered as irreducible elements, i.e. they cannot be expressed as
the product of several factorized terms.
In particular, the ghost numbers $G=(0,1)$ correspond to
an invariant gravitational Lagrangian and to an anomaly, respectively.

Thanks to the operator $\d$ and to the algebraic relations
\equ{DEC}-\equ{EXCOM}, in order to find a solution of the
ladder \equ{TOWER} it is sufficient to solve only the last equation
for the zero form $\O^{G+N}_{0}$.
It is easy to check that, once a non-trivial solution for $\O^{G+N}_{0}$
is known, the higher cocycles $\O^{G+N-q}_{q}$, $(q=1,...,N)$ are
obtained by repeated applications of the operator $\d$ on
$\O^{G+N}_{0}$, i.e.
\be
\label{HICO}
\O^{G+N-q}_{q}=\frac{\d^{q}}{q!}\O^{G+N}_{0}~~~,~~~q=1,...,N~~~,
{}~~~G=(0,1)~.
\ee
Therefore, the solution of the last equation of the tower
\equ{TOWER} is reduced to a problem of {\it local} BRST cohomology instead of a
modulo-$d$ one.
It is well-known indeed that, once a particular solution of the
descent equations \equ{TOWER} has been obtained, i.e. eq.\equ{HICO},
the search of the most general solution becomes essentially a problem
of local BRST cohomology.

The complete general solutions of the local cohomological problem
\be
s\O^{G+N}_{0}=0
\ee
are not yet obtained~\cite{next}, but it is rather simple to discuss some
interesting examples. This will be done in the next section.


\section{Some examples}

This section is devoted to apply the previous algebraic setup
and to discuss some explicit examples.
Especially, we draw our attention to the cohomological origin of the
cosmological constant, the Einstein Lagrangians, and the generalized
curvature Lagrangians.
In addition, we discuss Lagrangians with torsion,
Chern-Simons terms and anomalies.
In a last step we investigate the scalar field Lagrangians in the
presence of gravity, the Weyl anomalies, and the
Weyl invariant scalar field Lagrangians.
The analysis will be carried out for any space-time dimension, i.e.
the Lorentz group will be assumed to be $SO(N)$ with $N$ arbitrary.

Remark that in the following Sections 5.1-5.6 we will discuss
the gravitational case with torsion but without Weyl symmetry.
Therefore, in this sections one has to use the BRST transformations
summarized in Section 4.2.
In Sections 5.7 and 5.8 also Weyl symmetry is included, and there
one has to use the full BRST transformations of Section 3.3.


\subsection{The cosmological constant}

The simplest local BRST invariant polynomial which one can define is
\be
\label{COSMOZERO}
\O^{N}_{0}=\frac{(-1)^{N}}{N!}\ve_{a_{1}a_{2}.....a_{N}}
\h^{a_{1}}\h^{a_{2}}.....\h^{a_{N}} \ ,
\ee
with $\ve_{a_{1}a_{2}.....a_{N}}$ the totally antisymmetric invariant
tensor of $SO(N)$.
Taking into account that in a $N$-dimensional space-time the product
of $(N+1)$ ghost fields $\h^{a}$ automatically vanishes, it is easily
checked that $\O^{N}_{0}$ identifies a cohomology class of
the BRST operator, i.e.
\be
s\O^{N}_{0}=0~~~,~~~\O^{N}_{0} \not= s\widehat{\O}^{N-1}_{0} \ .
\ee
For the case $G=0$ the zero form cocycle \equ{COSMOZERO} corresponds to
the invariant Lagrangian $\O^{0}_{N}$
\be
\label{COSMOCONST}
\O_{N}^{0}=\frac{\d^{N}}{N!}\O^{N}_{0}
=\frac{1}{N!}\ve_{a_{1}a_{2}.....a_{N}}
e^{a_{1}}e^{a_{2}}.....e^{a_{N}} \ ,
\ee
which is easily identified with the $SO(N)$ cosmological
constant. One sees thus that the cohomological origin of the
cosmological constant \equ{COSMOCONST} relies on the cocycles
\equ{COSMOZERO}. With the help of Appendix B one can rewrite
eq.\equ{COSMOCONST} to the more familiar form
\be
\O_{N}^{0}=e^{1}.....e^{N}=e~d^{N}\!x \ .
\ee


\subsection{Einstein Lagrangians}

In this case, using the zero form curvature $R^{ab}_{~~mn}$, one gets
for the cocycle $\O^{N}_{0}$ $(N>2)$:
\be
\label{EINSTEINZERO}
\O^{N}_{0}=\frac{1}{2}\frac{(-1)^{N}}{(N-2)!}\ve_{a_{1}a_{2}.....a_{N}}
R^{a_{1}a_{2}}_{~~~~mn}\h^{m}\h^{n}\h^{a_{3}}.....\h^{a_{N}} \ ,
\ee
to which it corresponds the term
\bea
\label{EINSTEINLAGR}
\O^{0}_{N}\=\frac{\d^{N}}{N!}\O^{N}_{0}\non
\=\frac{1}{2}\frac{1}{(N-2)!}\ve_{a_{1}a_{2}.....a_{N}}
R^{a_{1}a_{2}}_{~~~~mn}e^{m}e^{n}e^{a_{3}}.....e^{a_{N}}\non
\=\frac{1}{(N-2)!}\ve_{a_{1}a_{2}.....a_{N}}
R^{a_{1}a_{2}}e^{a_{3}}.....e^{a_{N}} \ .
\eea
Expression \equ{EINSTEINLAGR} is nothing but the Einstein Lagrangian
for the case of $SO(N)$.
Using the result given in the Appendix B one gets
\bea
\O^{0}_{N}\=\frac{1}{2}\frac{1}{(N-2)!}\ve_{a_{1}a_{2}.....a_{N}}
R^{a_{1}a_{2}}_{~~~~mn}\ve^{mna_{3}.....a_{N}}e^{1}.....e^{N} \non
\=\frac{1}{2}eR^{a_{1}a_{2}}_{~~~~mn}(\d^{m}_{a_{1}}\d^{n}_{a_{2}}-
\d^{n}_{a_{1}}\d^{m}_{a_{2}})~d^{N}\!x \non
\=eR^{mn}_{~~~mn}~d^{N}\!x=eR~d^{N}\!x \ .
\eea
Notice also that for the case of $SO(2)$ the
zero form cocycle $\O^{2}_{0}$
\be
\O^{2}_{0}=\frac{1}{2}\ve_{ab}R^{ab}_{~~mn}\h^{m}\h^{n}
\ee
turns out to be BRST-exact:
\be
\O^{2}_{0}=-s(\ve_{ab}\o^{ab}_{~~m}\h^{m}+\ve_{ab}\th^{ab}) \ .
\ee
As it is well-known, this implies that the two dimensional
Einstein Lagrangian
\be
\O^{0}_{2}=\ve_{ab}R^{ab}
\ee
is $d$-exact, i.e.
\be
\O^{0}_{2}=d(\ve_{ab}\o^{ab}) \ .
\ee


\subsection{Generalized curvature Lagrangians}

A straightforward generalization of the Einstein Lagrangians
\equ{EINSTEINLAGR} is to replace any pair of
vielbeins with the two form $R^{ab}$.
Therefore, one gets another set of
gravitational Lagrangians containing higher powers of the
Riemann tensor.

To give an example, let us consider the zero form cocycle
\bea
\O^{2N}_{0}\=\frac{1}{(2N)!}\frac{1}{2^{N}}
(\ve_{a_{1}a_{2}a_{3}a_{4}.....a_{(2N-1)}a_{(2N)}}
R^{a_{1}a_{2}}_{~~~~b_{1}b_{2}}R^{a_{3}a_{4}}_{~~~~b_{3}b_{4}}
.....R^{a_{(2N-1)}a_{(2N)}}_{~~~~~~~~~~~~b_{(2N-1)}b_{(2N)}})\non
&\times&\!\!\!(\h^{b_{1}}\h^{b_{2}}\h^{b_{3}}\h^{b_{4}}.....
\h^{b_{(2N-1)}}\h^{b_{(2N)}}) \ .
\eea
Using eq.\equ{HICO}, for the corresponding invariant Lagrangian
one gets
\bea
\O^{0}_{2N}\=\frac{\d^{(2N)}}{(2N)!}\O^{2N}_{0}\non
\=\frac{1}{(2N)!}\frac{1}{2^{N}}
(\ve_{a_{1}a_{2}a_{3}a_{4}.....a_{(2N-1)}a_{(2N)}}
R^{a_{1}a_{2}}_{~~~~b_{1}b_{2}}R^{a_{3}a_{4}}_{~~~~b_{3}b_{4}}
.....R^{a_{(2N-1)}a_{(2N)}}_{~~~~~~~~~~~~b_{(2N-1)}b_{(2N)}})\non
&\times&\!\!\!(e^{b_{1}}e^{b_{2}}e^{b_{3}}e^{b_{4}}.....
e^{b_{(2N-1)}}e^{b_{(2N)}})\non
\=\frac{1}{(2N)!}
\ve_{a_{1}a_{2}a_{3}a_{4}.....a_{(2N-1)}a_{(2N)}}
R^{a_{1}a_{2}}R^{a_{3}a_{4}}.....R^{a_{(2N-1)}a_{(2N)}} \ .
\eea
As an explicit example we analyze the case of $SO(4)$ with the
cocycle
\bea
\O^{0}_{4}\=\frac{1}{4!}\frac{1}{4}
\ve_{a_{1}a_{2}a_{3}a_{4}}
R^{a_{1}a_{2}}_{~~~~b_{1}b_{2}}R^{a_{3}a_{4}}_{~~~~b_{3}b_{4}}
e^{b_{1}}e^{b_{2}}e^{b_{3}}e^{b_{4}} \non
\=\frac{1}{4!}\frac{1}{4}
\ve_{a_{1}a_{2}a_{3}a_{4}}\ve^{b_{1}b_{2}b_{3}b_{4}}
R^{a_{1}a_{2}}_{~~~~b_{1}b_{2}}R^{a_{3}a_{4}}_{~~~~b_{3}b_{4}}
{}~e~d^{4}\!x \non
\=\frac{1}{4!}(R^{ab}_{~~cd}R^{cd}_{~~ab}+4R^{ab}_{~~bd}R^{cd}_{~~ca}
+R^{ab}_{~~ab}R^{cd}_{~~cd})~e~d^{4}\!x \non
\=\frac{1}{4!}e(R^{ab}_{~~cd}R^{cd}_{~~ab}
-4R^{a}_{~d}R^{d}_{~a}+R^{2})~d^{4}\!x \ .
\eea
Above expression is nothing else but the Euler density~\cite{tonin3}.


\subsection{Lagrangians with torsion}

It is known that,
for special values of the space-time dimension
$N$, i.e. $N=(4M-1)$ with $M\ge1$, there is the possibility of defining
non-trivial invariant Lagrangians which explicitly contain the
torsion~\cite{baulieu}.
\newline
Let us begin by considering first the simpler case of $SO(3)$ $(M=1)$.
By making use of the zero form $T^{a}_{mn}$, one has for the cocycle
$\O^{3}_{0}$ the following expression\footnote{Tangent space indices are
rised and lowered with the flat metric $g_{ab}$, $\h_{a}=g_{ab}\h^{b}$.}
\be
\O^{3}_{0}=\frac{1}{2}T^{a}_{mn}\h^{m}\h^{n}\h_{a} \ ,
\ee
from which one gets the three dimensional torsion Lagrangian
\bea
\O^{0}_{3}=\frac{\d^{3}}{3!}\O^{3}_{0}
\=-\frac{1}{2}T^{a}_{mn}e^{m}e^{n}e_{a}
=-T^{a}e_{a} \ .
\eea
We remark that this term, known also as the
{\it translational Chern-Simons term}, has been already
discussed by several authors~\cite{baekler} and gives rise to
interesting gravitational models.
As shown by~\cite{baekler} it can be naturally included together
with the three dimensional topological Chern-Simons
term of Deser-Jackiw-Templeton~\cite{djt} and the cosmological constant
into the Einstein action. The resulting model is
characterized by the presence of a massive graviton moving in a
space of constant curvature.

Generalizing to the case of $SO(4M-1)$ with $(M>1)$, one finds
\bea
\O^{4M-1}_{0}\=\frac{1}{2^{(2M-1)}}(T_{k~\!\!m_{1}m_{2}}
R^{k}_{~a_{1}m_{3}m_{4}}R^{a_{1}}_{~~a_{2}m_{5}m_{6}}....
R^{a_{(2M-3)}}_{~~~~~~~a_{(2M-2)}m_{(4M-3)}m_{(4M-2)}})\non
&\times&\!\!\!(\h^{m_{1}}\h^{m_{2}}....\h^{m_{(4M-3)}}\h^{m_{(4M-2)}}
\h^{a_{(2M-2)}}) \ ,
\eea
which yields the following Lagrangians including also torsion
\bea
\O^{0}_{4M-1}\=-\frac{1}{2^{(2M-1)}}(T_{k~\!\!m_{1}m_{2}}
R^{k}_{~a_{1}m_{3}m_{4}}R^{a_{1}}_{~~a_{2}m_{5}m_{6}}....
R^{a_{(2M-3)}}_{~~~~~~~a_{(2M-2)}m_{(4M-3)}m_{(4M-2)}})\non
&\times&\!\!\!(e^{m_{1}}e^{m_{2}}....e^{m_{(4M-3)}}e^{m_{(4M-2)}}
e^{a_{(2M-2)}})\non
\=-T_{k}R^{k}_{~a_{1}}R^{a_{1}}_{~~a_{2}}....
R^{a_{(2M-3)}}_{~~~~~~~a_{(2M-2)}}
e^{a_{(2M-2)}} \ .
\eea

Let us also mention the possibility of defining invariant
Lagrangians with torsion terms
which are polynomial in $T^{a}_{mn}$. These Lagrangians exist in any
space-time dimension and are easily obtained from the
$SO(N)$ zero form cocycle
\be
\label{POLTORS}
\O^{N}_{0}=\frac{(-1)^{N}}{N!}\ve_{a_{1}a_{2}.....a_{N}}
\h^{a_{1}}\h^{a_{2}}.....\h^{a_{N}} \PP(T) \ ,
\ee
where $\PP(T)$ is a scalar polynomial in the torsion as,
for instance~\cite{kumm} (see also~\cite{sez} for generalization),
\be
  \PP(T) = T^{a}_{mn} T_{a}^{mn} \ .
\label{PTEXAMPLE}
\ee
The corresponding invariant Lagrangians containing only torsion
are then given by
\be
\label{PTLAGRANG}
\O_{N}^{0}=\frac{\d^{N}}{N!}\O^{N}_{0}
=\frac{1}{N!}\ve_{a_{1}a_{2}.....a_{N}}
e^{a_{1}}e^{a_{2}}.....e^{a_{N}}\PP(T) \ .
\ee


\subsection{Chern-Simons terms and anomalies}

For what concerns the Chern-Simons terms and the Lorentz and
diffeomorphism anomalies we recall that, as mentioned
in the introduction, a systematic analysis based on the decomposition
\equ{DEC} has been recently carried out by~\cite{werneck}.

Let us remark, however, that the decomposition found in~\cite{werneck}
gives
rise to a commutation relation between the operators $\d$ and $d$ which,
contrary to the present case (see eq.\equ{EXCOM}), does not vanish
(see also Section 4.1).
This implies the existence of a further operator $\GG$ of degree one
which has to be taken into account in order to solve the
ladder \equ{TOWER}.

Actually, the existence of the operator $\GG$ relies on the fact that
the decomposition of the exterior differential $d$ found
in~\cite{werneck}
does not take into account the explicit presence of the vielbein
$e^{a}$ and of the torsion $T^{a}$. It holds for a functional space
whose basic elements are built up only with the Lorentz connection
$\o^{a}_{~b}$ and the Riemann tensor $R^{a}_{~b}$, this choice
being sufficient to characterize all known Lorentz anomalies and
related second family diffeomorphism cocycles~\cite{tonin1,tonin2}.

It is remarkable then to observe that the algebra between $s$, $\d$,
and $d$ gets simpler only when the vielbein $e^{a}$ and the torsion
$T^{a}$ are naturally present.
Let us emphasize indeed that the particular elementary form of the
operator $\d$ in eq.\equ{DECETA} is due to the use of the
tangent space ghost $\h^{a}$ whose introduction requires explicitly
the presence of the vielbein $e^{a}$.

For the sake of clarity and to make contact with the results obtained
in~\cite{werneck}, let us discuss in details the construction of the
$SO(3)$ Chern-Simons term. In this case the tower \equ{TOWER} takes
the form
\bea
&&s\O^{0}_{3}+d\O^{1}_{2}=0 \ ,\non
&&s\O^{1}_{2}+d\O^{2}_{1}=0 \ ,\non
&&s\O^{2}_{1}+d\O^{3}_{0}=0 \ ,\non
&&s\O^{3}_{0}=0 \ ,
\eea
where, according to eq.\equ{HICO},
\bea
&&\O^{2}_{1}=\d\O^{3}_{0} \ ,\non
&&\O^{1}_{2}=\frac{\d^{2}}{2!}\O^{3}_{0} \ ,\non
&&\O^{0}_{3}=\frac{\d^{3}}{3!}\O^{3}_{0} \ .
\eea
In order to find a solution for $\O^{3}_{0}$ we use the redefined
Lorentz ghost
\be
\widehat{\th}^{a}_{~b}=\o^{a}_{~bm}\h^{m}+\th^{a}_{~b} \ ,
\ee
which, from eq.\equ{DECETA}, transforms as
\be
\d\widehat{\th}^{a}_{~b}=-\o^{a}_{~b} \ .
\ee
For the cocycle $\O^{3}_{0}$ one gets then
\be
\O^{3}_{0}=\frac{1}{3}\widehat{\th}^{a}_{~b}
\widehat{\th}^{\hspace{0.04cm}b}_{~c}
\widehat{\th}^{c}_{~a}-\frac{1}{2}R^{a}_{~bmn}\h^{m}\h^{n}
\widehat{\th}^{\hspace{0.04cm}b}_{~a} \ ,
\ee
from which $\O^{2}_{1}$, $\O^{1}_{2}$, and $\O^{0}_{3}$ are
computed to be
\bea
\O^{2}_{1}=-\o^{a}_{~b}
\widehat{\th}^{\hspace{0.04cm}b}_{~c}
\widehat{\th}^{c}_{~a}
+R^{a}_{~bmn}e^{m}\h^{n}
\widehat{\th}^{\hspace{0.04cm}b}_{~a}
+\frac{1}{2}R^{a}_{~bmn}\h^{m}\h^{n}\o^{b}_{~a} \ ,
\eea
\bea
\label{CHERNTWO}
\O^{1}_{2}=\o^{a}_{~b}\o^{b}_{~c}\widehat{\th}^{c}_{~a}
-R^{a}_{~b}\widehat{\th}^{\hspace{0.04cm}b}_{~a}
-R^{a}_{~bmn}e^{m}\h^{n}\o^{b}_{~a} \ ,
\eea
\bea
\label{CHERNTHREE}
\O^{0}_{3}=R^{a}_{~b}\o^{b}_{~a}
-\frac{1}{3}\o^{a}_{~b}\o^{b}_{~c}\o^{c}_{~a} \ .
\eea
In particular, expression \equ{CHERNTHREE} gives the familiar $SO(3)$
Chern-Simons gravitational term. Finally, let us remark that the
cocycle $\O^{1}_{2}$ of eq.\equ{CHERNTWO}, when referred to $SO(2)$,
reduces to the expression
\be
\O^{1}_{2}=-(d\o^{a}_{~b})\th^{b}_{~a} \ ,
\ee
which directly gives the two dimensional Lorentz anomaly.
Analogous, for the zero form cocycle $\O^{5}_{0}$ in $SO(5)$ one gets
\bea
\O^{5}_{0}\=-\frac{1}{10}\widehat{\th}^{a}_{~b}
\widehat{\th}^{\hspace{0.04cm}b}_{~c}
\widehat{\th}^{\hspace{0.04cm}c}_{~d}
\widehat{\th}^{\hspace{0.04cm}d}_{~e}
\widehat{\th}^{\hspace{0.04cm}e}_{~a}
+\frac{1}{4}R^{a}_{~bmn}\h^{m}\h^{n}
\widehat{\th}^{\hspace{0.04cm}b}_{~c}
\widehat{\th}^{\hspace{0.04cm}c}_{~d}
\widehat{\th}^{\hspace{0.04cm}d}_{~a} \non
\-\frac{1}{4}R^{a}_{~bmn}\h^{m}\h^{n}R^{b}_{~ckl}\h^{k}\h^{l}
\widehat{\th}^{\hspace{0.04cm}c}_{~a} \ ,
\eea
which leads to the five dimensional Chern-Simons term
\bea
\O^{0}_{5}\=\frac{1}{10}\o^{a}_{~b}\o^{b}_{~c}\o^{c}_{~d}
\o^{d}_{~e}\o^{e}_{~a}
-\frac{1}{4}R^{a}_{~bmn}e^{m}e^{n}
\o^{b}_{~c}\o^{c}_{~d}\o^{d}_{~a} \non
\+\frac{1}{4}R^{a}_{~bmn}e^{m}e^{n}R^{b}_{~ckl}e^{k}e^{l}
\o^{c}_{~a} \non
\=\frac{1}{10}\o^{a}_{~b}\o^{b}_{~c}\o^{c}_{~d}
\o^{d}_{~e}\o^{e}_{~a}
-\frac{1}{2}R^{a}_{~b}\o^{b}_{~c}\o^{c}_{~d}\o^{d}_{~a}
+R^{a}_{~b}R^{b}_{~c}\o^{c}_{~a} \ .
\eea


\subsection{Scalar field Lagrangians}

In order to couple a massless scalar field to gravity
without Weyl symmetry one now
considers the following zero form cocycle $\O^{N}_{0}$
in a $N$-dimensional space-time
\be
\label{SCALAR-ZERO-0}
\O^{N}_{0}=\frac{1}{2}\frac{1}{N!}\ve_{a_{1}a_{2}.....a_{N}}
\h^{a_{1}}\h^{a_{2}}.....\h^{a_{N}}(D_{m}\ph)(D^{m}\ph) \ ,
\ee
where the covariant derivative reduces to the ordinary one
\be
\label{SCALAR-ZERO}
\O^{N}_{0}=\frac{1}{2}\frac{1}{N!}\ve_{a_{1}a_{2}.....a_{N}}
\h^{a_{1}}\h^{a_{2}}.....\h^{a_{N}}(\6_{m}\ph)(\6^{m}\ph) \ .
\ee
Taking into account the truncated BRST transformation
\be
s(\6_{m}\ph)=-\th^{k}_{~m}(\6_{k}\ph)-\h^{k}\6_{k}(\6_{m}\ph) \ ,
\ee
it can be easily
checked that $\O^{N}_{0}$ is BRST invariant, i.e.
\be
s\O^{N}_{0}=0 \ .
\ee
For the case $G=0$ one gets the invariant Lagrangian $\O^{0}_{N}$
\be
\label{SCALAR-LAGRANGIAN}
\O_{N}^{0}=\frac{\d^{N}}{N!}\O^{N}_{0}
=\frac{1}{2}\frac{(-1)^{N}}{N!}\ve_{a_{1}a_{2}.....a_{N}}
e^{a_{1}}e^{a_{2}}.....e^{a_{N}}(\6_{m}\ph)(\6^{m}\ph) \ ,
\ee
which is easily recognized to coincide with the $SO(N)$
scalar field Lagrangian
\bea
\O_{N}^{0}\=\frac{1}{2}d^{N}\!x~e(\6_{m}\ph)(\6^{m}\ph) \non
\=\frac{1}{2}d^{N}\!x~\sqrt{g}(\6_{\mu}\ph)(\6^{\mu}\ph) \ .
\eea
Notice that above scalar field Lagrangian is invariant under
diffeomorphisms and local Lorentz rotations, but not invariant
under Weyl transformations.
This case will be studied in Section 5.8.


\subsection{Weyl anomalies}

In order to incorporate also the Weyl symmetry we use
from now on the full BRST operator $s$, defined in
Section 3.3. For a better understanding of the matter let us begin
by discussing the simplest case, the Weyl anomaly in $SO(2)$.
According to \equ{TOWER} we have therefore to solve the tower
\bea
&&s\O^{1}_{2}+d\O^{2}_{1}=0\non
&&s\O^{2}_{1}+d\O^{3}_{0}=0\non
&&s\O^{3}_{0}=0 \ ,
\eea
where $\O^{1}_{2}$ is denoting the corresponding anomaly.
For the zero form cocycle $\O^{3}_{0}$ one has
\be
\O^{3}_{0}=\frac{1}{2}\ve_{ab}\h^{a}\h^{b}\s R \ ,
\ee
with $R$ as the Riemann scalar \equ{RICCI-SCALAR}.
One can easy verify with
the BRST transformation of the Riemann scalar $R$, given by
\be
sR=-\h^{k}\6_{k}R-2\s R \ ,
\ee
that above cocycle is BRST invariant, i.e.
\be
s\O^{3}_{0}=0 \ .
\ee
By using eq.\equ{HICO} one gets the well-known two dimensional
Weyl anomaly
\bea
\O^{1}_{2}\=\frac{\d^{2}}{2!}\O^{3}_{0} \non
\=\frac{1}{2}\ve_{ab}e^{a}e^{b}\s R \ ,
\eea
which can be rewritten to the more familiar form (see Appendix B)
\be
\O^{1}_{2}=e\s R~d^{2}\!x \ .
\ee

The second example, which we will discuss now, is the four dimensional
Weyl anomaly in $SO(4)$.
One possible zero form cocycle $\O^{5}_{0}$ is given by
\bea
\O^{5}_{0}=\frac{1}{4!}\ve_{abcd}\h^{a}\h^{b}\h^{c}\h^{d}\s R^{2} \ ,
\eea
which is BRST invariant, i.e.
\be
s\O^{5}_{0}=0 \ .
\ee
This leads to the anomaly
\bea
\O^{1}_{4}\=\frac{\d^{4}}{4!}\O^{5}_{0} \non
\=\frac{1}{4!}\ve_{abcd}e^{a}e^{b}e^{c}e^{d}\s R^{2} \non
\=e\s R^{2}~d^{4}\!x \ .
\eea
Two further possible zero form cocycles are given by
\bea
\label{XX}
\O^{5}_{0}=\frac{1}{4!}\ve_{abcd}\h^{a}\h^{b}\h^{c}\h^{d}\s R_{mn}R^{mn} \ ,
\eea
and
\bea
\label{XXX}
\O^{5}_{0}=\frac{1}{4!}\ve_{abcd}\h^{a}\h^{b}\h^{c}\h^{d}\s R_{mnkl}R^{mnkl} \
,
\eea
where the zero forms $R_{mn}$ and $R_{mnkl}$ denoting the Ricci tensor
and the Riemann tensor with indices in the tangent space. The cocycles
\equ{XX} and \equ{XXX} are again BRST invariant:
\be
s\O^{5}_{0}=0 \ .
\ee
{}From \equ{XX} and \equ{XXX} one gets for the corresponding anomalies
\bea
\O^{1}_{4}\=\frac{\d^{4}}{4!}\O^{5}_{0} \non
\=\frac{1}{4!}\ve_{abcd}e^{a}e^{b}e^{c}e^{d}\s R_{mn}R^{mn} \non
\=e\s R_{mn}R^{mn}~d^{4}\!x \ ,
\eea
and
\bea
\O^{1}_{4}\=\frac{\d^{4}}{4!}\O^{5}_{0} \non
\=\frac{1}{4!}\ve_{abcd}e^{a}e^{b}e^{c}e^{d}\s R_{mnkl}R^{mnkl} \non
\=e\s R_{mnkl}R^{mnkl}~d^{4}\!x \ .
\eea


{}From the variety of all possible cocycles in higher dimensions we quote
only the simplest example for a zero form cocycle in $SO(2N)$
which has the following form:
\bea
\label{2N-DIM-ZEROCOCYCLE}
\O^{2N+1}_{0}=\frac{1}{(2N)!}\ve_{a_{1}a_{2}.....a_{2N}}
\h^{a_{1}}\h^{a_{2}}.....\h^{a_{2N}}\s R^{N} \ .
\eea
Of course, it can be easily checked that all these cocycles are
BRST invariant, i.e.
\be
s\O^{2N+1}_{0}=0 \ .
\ee
The corresponding $2N$-dimensional Weyl anomalies are given by
\bea
\O^{1}_{2N}\=\frac{\d^{2N}}{(2N)!}\O^{2N+1}_{0} \non
\=\frac{1}{(2N)!}\ve_{a_{1}a_{2}.....a_{2N}}
e^{a_{1}}e^{a_{2}}.....e^{a_{2N}}\s R^{N} \non
\=e\s R^{N}~d^{2N}\!x \ .
\eea



\subsection{Weyl invariant scalar field Lagrangians}

In order to couple a massless scalar field to gravity
including also Weyl symmetry we
consider the following zero form cocycle $\O^{N}_{0}$
in a $N$-dimensional space-time
\be
\label{WEYL-SCALAR-ZERO-0}
\O^{N}_{0}=\frac{1}{2}\frac{1}{N!}\ve_{a_{1}a_{2}.....a_{N}}
\h^{a_{1}}\h^{a_{2}}.....\h^{a_{N}}(D_{m}\ph)(D^{m}\ph) \ ,
\ee
where now the covariant derivative reduces to the Weyl covariant
derivative \equ{WEYL-COVD}
\be
\label{WEYL-COVD-0}
\nabla_{m}\ph=\6_{m}\ph-\frac{N-2}{2}A_{m}\ph \ ,
\ee
and one gets
\be
\label{WEYL-SCALAR-ZERO}
\O^{N}_{0}=\frac{1}{2}\frac{1}{N!}\ve_{a_{1}a_{2}.....a_{N}}
\h^{a_{1}}\h^{a_{2}}.....\h^{a_{N}}(\nabla_{m}\ph)(\nabla^{m}\ph) \ .
\ee
Notice, that from eq.\equ{WEYL-COVD-NABLA} follows
\be
s_{w}(\nabla_{\mu}\ph)=-\frac{N-2}{2}\s (\nabla_{\mu}\ph) \ ,
\ee
where $s_{w}$ denotes only the Weyl symmetry part of the full BRST operator.
Therefore, one obtains for \equ{WEYL-COVD-0} the Weyl-BRST transformation
\be
s_{w}(\nabla_{m}\ph)=s_{w}(E_{m}^{\mu}\nabla_{\mu}\ph)
=-\frac{N}{2}\s (\nabla_{m}\ph) \ .
\ee
Taking into account the full BRST transformation (including Weyl symmetry)
\be
s(\nabla_{m}\ph)=-\th^{k}_{~m}(\nabla_{k}\ph)-\h^{k}\6_{k}(\nabla_{m}\ph)
-\frac{N}{2} \s (\nabla_{m}\ph) \ ,
\ee
it can be easily
checked that $\O^{N}_{0}$ is BRST invariant, i.e.
\be
s\O^{N}_{0}=0 \ .
\ee
For the case $G=0$ one gets the Weyl invariant Lagrangian $\O^{0}_{N}$
\be
\label{WEYL-SCALAR-LAGRANGIAN}
\O_{N}^{0}=\frac{\d^{N}}{N!}\O^{N}_{0}
=\frac{1}{2}\frac{(-1)^{N}}{N!}\ve_{a_{1}a_{2}.....a_{N}}
e^{a_{1}}e^{a_{2}}.....e^{a_{N}}(\nabla_{m}\ph)(\nabla^{m}\ph) \ ,
\ee
which is easily recognized to coincide with the $SO(N)$
Weyl invariant scalar field Lagrangian
\bea
\O_{N}^{0}\=\frac{1}{2}d^{N}\!x~e(\nabla_{m}\ph)(\nabla^{m}\ph) \non
\=\frac{1}{2}d^{N}\!x~\sqrt{g}(\nabla_{\mu}\ph)(\nabla^{\mu}\ph) \ .
\eea
Notice that above scalar field Lagrangian is invariant under
diffeomorphisms, local Lorentz rotations, and also invariant
under Weyl transformations.


\section{The geometrical meaning of the operator $\d$}

Having discussed the role of the operator $\d$ in finding explicit
solutions of the descent equations \equ{TOWER}, let us turn now to
the study of its geometrical meaning.
As we shall see, this operator turns out to possess a quite simple
geometrical interpretation which will reveal an unexpected and so far
unnoticed elementary structure of the ladder \equ{TOWER}.

Let us begin by observing that all the cocycles $\O^{G+N-p}_{p}$
$(p=0,...,N)$ entering the descent equations \equ{TOWER} are of the
same degree (i.e. $(G+N)$), the latter being given by the sum of
the ghost number and of the form degree.

We can collect then, following~\cite{tataru}, all the $\O^{G+N-p}_{p}$
into a unique cocycle $\widehat{\O}$ of degree $(G+N)$ defined as
\be
\widehat{\O}=\sum_{p=0}^{N}\O^{G+N-p}_{p} \ .
\ee
This expression, using eq.\equ{HICO}, becomes
\be
\label{SUM}
\widehat{\O}=\sum_{p=0}^{N}\frac{\d^{p}}{p!}\O^{G+N}_{0} \ ,
\ee
where the cocycle $\O^{G+N}_{0}$, according to its zero form degree,
depends only on the set of zero form variables
$(\o^{a}_{~bm}, R^{a}_{~bmn}, T^{a}_{mn}, \th^{a}_{~b}, \h^{a})$
and their tangent space derivatives $\6_{m}$.
Taking into account that under the action of the operator $\d$ the
form degree and the ghost number are respectively raised and lowered
by one unit and that in a space-time of dimension $N$ a $(N+1)$-form
identically vanishes, it follows that eq.\equ{SUM} can be rewritten
in a more suggestive way as
\be
\label{EXPONENT}
\widehat{\O}=e^{\d}\O^{G+N}_{0}(\h^{a}, \th^{a}_{~b}, \o^{a}_{~bm},
R^{a}_{~bmn}, T^{a}_{mn}) \ .
\ee
Let us make now the following elementary but important remark.
As one can see from eq.\equ{DECETA}, the operator $\d$ acts as a
translation on the ghost $\h^{a}$ with an amount given by $(-e^{a})$.
Therefor $e^{\d}$ has the simple effect of shifting $\h^{a}$ into
$(\h^{a}-e^{a})$. This implies that the cocycle \equ{EXPONENT} takes
the form
\be
\label{SHIFT}
\widehat{\O}=\O^{G+N}_{0}(\h^{a}-e^{a}, \th^{a}_{~b}, \o^{a}_{~bm},
R^{a}_{~bmn}, T^{a}_{mn}) \ .
\ee
This formula collects in a very elegant and simple expression the
solution of the descent equations \equ{TOWER}.

In particular, it states the important result that:

\begin{quote}
 {\it To find a non-trivial solution of the ladder} \equ{TOWER}
{\it it
is sufficient to replace the variable $\h^{a}$ with $(\h^{a}-e^{a})$
in the zero form cocycle $\O^{G+N}_{0}$ which belongs to the local
cohomology of the BRST operator $s$. The expansion
of $\O^{G+N}_{0}$($\h^{a}-e^{a}$, $\th^{a}_{~b}$, $\o^{a}_{~bm}$,
$R^{a}_{~bmn}$, $T^{a}_{mn}$) in powers of the one form vielbein
$e^{a}$ yields then all the searched cocycles $\O^{G+N-p}_{p}$.}
\end{quote}

It is a simple exercise to check now that all the invariant
Lagrangians and Chern-Simons terms computed in the previous section
are indeed recovered by simply expanding the corresponding zero form
cocycles $\O^{G+N}_{0}$ taken as functions of $(\h^{a}-e^{a})$.

Let us conclude by remarking that, up to our knowledge, expression
\equ{SHIFT} represents a deeper understanding of the algebraic
properties of the gravitational ladder \equ{TOWER} and of the role
played by the vielbein $e^{a}$ and the associated ghost $\h^{a}$.


\section{Conclusion}

The algebraic structure of gravity with torsion in the presence of
Weyl symmetry has been analyzed
in the context of the Maurer-Cartan horizontality formalism by
introducing an operator $\delta$ which allows
to decompose the exterior space-time derivative as a BRST commutator.
Such a decomposition gives a simple and elegant way of solving
the Wess-Zumino consistency condition corresponding to invariant
Lagrangians and anomalies. The same technique can be
applied to the study of the gravitational coupling of Yang-Mills
gauge theories as well as to the characterization of
the Weyl anomalies~\cite{next}.


\section*{Acknowledgements}

We are especially grateful to A. Brandhuber, M. Werneck
de Oliveira and S.P. Sorella for all the time we spent together
in understanding and clarifying several aspects of this manuscript.


\section*{Appendices:}

Appendix A is devoted to demonstrate the computation of some commutators
involving the tangent space derivative $\6_{a}$ introduced in
Section 3. In the Appendix B one finds the definition of the
determinant of the vielbein in connection with the $\ve$ tensor.


\section*{A~~~Commutator relations}

\setcounter{equation}{0}
\renewcommand{\theequation}{A.\arabic{equation}}

In order to find the commutator of two tangent space derivatives
$\6_{a}$, we make use of the fact that the usual space-time derivatives
$\6_{\mu}$ have a vanishing commutator:
\be
[\6_{\mu},\6_{\nu}]=0 \ .
\ee
{}From
\be
\6_{\mu}=e^{m}_{\mu}\6_{m}
\ee
one gets
\bea
[\6_{\mu},\6_{\nu}]=0 \= [e^{m}_{\mu}\6_{m},e^{n}_{\nu}\6_{n}]\non
\= e^{m}_{\mu}e^{n}_{\nu}[\6_{m},\6_{n}]
+e^{m}_{\mu}(\6_{m}e^{n}_{\nu})\6_{n}
-e^{n}_{\nu}(\6_{n}e^{m}_{\mu})\6_{m}\non
\=e^{m}_{\mu}e^{n}_{\nu}[\6_{m},\6_{n}]
+(\6_{\mu}e^{k}_{\nu}-\6_{\nu}e^{k}_{\mu})\6_{k}\non
\=e^{m}_{\mu}e^{n}_{\nu}[\6_{m},\6_{n}]
+(T^{k}_{\mu\nu}-\o^{k}_{~n\mu}e^{n}_{\nu}
+\o^{k}_{~m\nu}e^{m}_{\mu}
-A_{\mu}e^{k}_{\nu}+A_{\nu}e^{k}_{\mu})\6_{k}\non
\=e^{m}_{\mu}e^{n}_{\nu}(T^{k}_{mn}+\o^{k}_{~mn}-\o^{k}_{~nm}
-A_{m}\d^{k}_{n}+A_{n}\d^{k}_{m})\6_{k} \non
\+e^{m}_{\mu}e^{n}_{\nu}[\6_{m},\6_{n}] \ ,
\eea
so that
\be
[\6_{m},\6_{n}]=-(T^{k}_{mn}+\o^{k}_{~mn}-\o^{k}_{~nm}
-A_{m}\d^{k}_{n}+A_{n}\d^{k}_{m})\6_{k} \ .
\ee
\newline
For the commutator of $d$ and $\6_{m}$ we get
\bea
[d,\6_{m}]\=[e^{n}\6_{n},\6_{m}]\non
\=-(\6_{m}e^{k})\6_{k}-e^{n}[\6_{m},\6_{n}]\non
\=-(\6_{m}e^{k})\6_{k}+e^{n}(T^{k}_{mn}+\o^{k}_{~mn}
-\o^{k}_{~nm}-A_{m}\d^{k}_{n}+A_{n}\d^{k}_{m})\6_{k} \ ,
\eea
and one has therefore
\be
[d,\6_{m}]=(T^{k}_{mn}e^{n}+\o^{k}_{~mn}e^{n}
-\o^{k}_{~nm}e^{n}-A_{m}e^{k}+A_{n}e^{n}\d^{k}_{m}
-(\6_{m}e^{k}))\6_{k} \ .
\ee
Analogously, from
\be
[s,\6_{\mu}]=0
\ee
one easily finds
\bea
[s,\6_{m}]\=(\6_{m}\h^{k}-\th^{k}_{~m}-\s\d^{k}_{m})\6_{k}
+\h^{n}[\6_{m},\6_{n}]\non
\=(\6_{m}\h^{k}-\th^{k}_{~m}-T^{k}_{mn}\h^{n}
-\o^{k}_{~mn}\h^{n}+\o^{k}_{~nm}\h^{n} \non
\+A_{m}\h^{k}-A_{n}\h^{n}\d^{k}_{m}-\s\d^{k}_{m})\6_{k} \ .
\eea


\section*{B~~~Determinant of the vielbein and the $\ve$ tensor}

\setcounter{equation}{0}
\renewcommand{\theequation}{B.\arabic{equation}}

The definition of the determinant of the vielbein $e^{a}_{\mu}$
is given by
\bea
e\=det(e^{a}_{\mu})=\frac{1}{N!}\ve_{a_{1}a_{2}.....a_{N}}
\ve^{\mu_{1}\mu_{2}.....\mu_{N}}e^{a_{1}}_{\mu_{1}}e^{a_{2}}_{\mu_{2}}
.....e^{a_{N}}_{\mu_{N}} \ .
\eea
One can easily verify that the BRST transformation of $e$ reads
\be
se=-\6_{\l}(\x^{\l}e) \ .
\ee
For the volume element one has
\bea
e^{1}.....e^{N}\=\frac{1}{N!}\ve_{a_{1}.....a_{N}}
e^{a_{1}}.....e^{a_{N}}\non
\=\frac{1}{N!}\ve_{a_{1}.....a_{N}}e^{a_{1}}_{\mu_{1}}.....
e^{a_{N}}_{\mu_{N}}
dx^{\mu_{1}}.....dx^{\mu_{N}}\non
\=\frac{1}{N!}\ve_{a_{1}.....a_{N}}\ve^{\mu_{1}.....\mu_{N}}
e^{a_{1}}_{\mu_{1}}.....e^{a_{N}}_{\mu_{N}}
dx^{1}.....dx^{N}\non
\=ed^{N}\!x=\sqrt{g}d^{N}\!x \ ,
\eea
where $g$ denotes the determinant of the metric tensor $g_{\mu\nu}$
\be
g=det(g_{\mu\nu}) \ .
\ee
The $\ve$ tensor has the usual norm
\be
\ve_{a_{1}.....a_{N}}\ve^{a_{1}.....a_{N}}=N! \ ,
\ee
and obeys the following relation under partial contraction
of $(N-2)$ indices
\be
\ve_{a_{1}.....a_{N}}\ve^{mna_{3}.....a_{N}}
=(N-2)!(\d^{m}_{a_{1}}\d^{n}_{a_{2}}-\d^{n}_{a_{1}}\d^{m}_{a_{2}}) \ ,
\ee
and in general the contraction of two $\ve$ tensors is given by
the determinant of $\d$ tensors in the following way
\be
\ve_{a_{1}.....a_{N}}\ve^{b_{1}.....b_{N}}=
\left|
\begin{array}{cccc}
\d^{b_{1}}_{a_{1}} & \d^{b_{2}}_{a_{1}} & ..... & \d^{b_{N}}_{a_{1}}  \\
\d^{b_{1}}_{a_{2}} & \d^{b_{2}}_{a_{2}} & ..... & \d^{b_{N}}_{a_{2}}  \\
..... & ..... & ..... & .....  \\
\d^{b_{1}}_{a_{N}} & \d^{b_{2}}_{a_{N}} & ..... & \d^{b_{N}}_{a_{N}}
\end{array}
\right| \ .
\ee



\end{document}